\RequirePackage{fix-cm}
\documentclass[pdftexify,twocolumn,epjc3]{svjour3}          
\RequirePackage[T1]{fontenc}
\RequirePackage{color,soul}
\RequirePackage[figtopcap]{subfigure}
\RequirePackage{graphicx}
\RequirePackage{amssymb}
\RequirePackage{mathptmx}      
\RequirePackage{flushend}
\RequirePackage[numbers,sort&compress]{natbib}
\RequirePackage[colorlinks,citecolor=blue,urlcolor=blue,linkcolor=blue]{hyperref}
\smartqed  
\journalname{Eur. Phys. J. C}
\newcommand{\overcirc}[1]{{\mathop{#1}\limits^{\tiny{\circ}}}}
\newcommand{\dddot}[1]{\ {\mathop{#1}\limits^{{\tiny{\textrm{...}}}}}\ }
\begin{document}
\title{The Hidden Flat Like Universe}
\subtitle{Starobinsky-like inflation induced by $f(T)$ gravity}
\author{W. El Hanafy\thanksref{addr1,addr3,e1}
\and
G.G.L. Nashed\thanksref{addr1,addr2,addr3,e2}
}
\thankstext{e1}{e-mail: waleed.elhanafy@bue.edu.eg}
\thankstext{e2}{e-mail: nashed@bue.edu.eg}
\institute{Centre for theoretical physics, the British University in Egypt, 11837 - P.O. Box 43, Egypt\label{addr1}
          \and
          Mathematics Department, Faculty of Science, Ain Shams University, Cairo, Egypt\label{addr2}
          \and
          Egyptian Relativity Group, Egypt\label{addr3}
}

\date{Received: date / Accepted: date}
\maketitle
\begin{abstract}
We study a single fluid component in a flat like universe (FLU) governed by $f(T)$ gravity theories, where $T$ is the teleparallel torsion scalar. The FLU model, regardless the value of the spatial curvature $k$, identifies a special class of $f(T)$ gravity theories. Remarkably, the FLU $f(T)$ gravity does not reduce to teleparallel gravity theory. In large Hubble spacetime the theory is consistent with the inflationary universe scenario and respects the conservation principle. The equation of state (EoS) evolves similarly in all models $k=0, \pm 1$. We study the case when the torsion tensor is made of a scalar field, which enables to derive a quintessence potential from the obtained $f(T)$ gravity theory. The potential produces Starobinsky-like model naturally without using a conformal transformation, with higher orders continuously interpolate between Starobinsky and quadratic inflation models. The slow-roll analysis shows double solutions so that for a single value of the scalar tilt (spectral index) $n_{s}$ the theory can predict double tensor-to-scalar ratios $r$ of $E$-mode and $B$-mode polarizations.
\end{abstract}
\section{Introduction}\label{Sec1}
The general relativity (GR) theory explained the gravity as  spacetime curvature. This  description of gravitation has succeeded to confront astrophysical observations for a long time. It has predicted perfectly the perihelion shift of Mercury, time delay in the solar system. Even in the strong field regimes such as binary pulsars it has amazingly predicted their orbital decay due to gravitational radiation by the system. While it fails to predict the accelerating cosmic expansion which is evidenced by the astronomical observations of high-redshift Type Ia supernovae \cite{SN98}. The teleparallel equivalent of general relativity (TEGR) theory has provided an alternative description of Einstein's gravity. The theory constructed from vierbein (tetrad) fields\footnote{The Greek letters $\alpha,\beta,...$ denote the spacetime indices and the Latin ones $a,b,...$ denote Lorentz indices. Both run from $0$ to $3$.} $\{h^{a}{_\mu}\}$ instead of metric tensor fields $g_{\mu \nu}$. The metric space, however, can be constructed from the vierbein fields, so the Levi-Civita symmetric connection $\overcirc{\Gamma}{^{\alpha}}{_{\mu \nu}}$. It is, also, possible to construct Weitzenb\"{o}ck nonsymmetric connection $\Gamma^{\alpha}{_{\mu\nu}}$. The first connection has a non-vanishing curvature tensor $R^{(\overcirc{\Gamma})}_{\alpha \beta \mu \nu}\neq 0$ but a vanishing torsion tensor $T^{(\overcirc{\Gamma})}_{\alpha \mu \nu}=0$, while the later is characterized by a vanishing curvature tensor $R^{(\Gamma)}_{\alpha \beta \mu \nu}=0$ but a non-vanishing torsion tensor $T^{(\Gamma)}_{\alpha \mu \nu}\neq 0$. The combined picture could be encoded in the contracted Bianchi identity
$$R^{(\overcirc{\Gamma})}\equiv -T^{(\Gamma)}-2\nabla^{(\overcirc{\Gamma})}_{\alpha}T^{\nu\alpha}{_\nu},$$
where $R^{(\overcirc{\Gamma})}$ is the usual Ricci scalar, the scalar invariant $T^{(\Gamma)}$ is called the teleparallel torsion scalar (it will be briefly investigated in Section \ref{Sec2}  and the covariant derivative $\nabla^{(\overcirc{\Gamma})}$ is with respect to (w.r.t.) the Levi-Civita connection. The variation of the left hand side w.r.t. the metric tensor provides the GR field equations. Since the last term in the identity is a total derivative, it does not contribute to the field equations, and the variation of the right hand side w.r.t. the vierbein fields provides a set of field equations equivalent to the GR that is called TEGR. Although the two theories are quantitatively equivalent at their level of the field equations, they are qualitatively different at the level of their actions! Indeed, the total derivative term is scalar invariant under diffeomorphism but  not invariant under local Lorentz transformation (LLT). On the other hand, the Ricci scalar in Einstein-Hilbert action leads to a theory invariant under diffeomorphism as well as LLT. Consequently, the teleparallel torsion scalar $T$ is not invariant under LLT \cite{AP2013}. The absence of local Lorentz symmetry in TEGR \textit{action} will not be reflected in the field equations, so it does not worth to worry about it. However, the presence of the total derivative term is crucial when consider the $f(T)$ extension of TEGR \cite{1010.1041,1012.4039}.

The telleparallel geometry has received attentions in the last decade. However, this geometry has been used very earlier in 1920s to unify gravity and electromagnetism by Einstein \cite{E28}. After this trial the geometry has been developed \cite{M52, MM56,M64}. Later, successful extensions to Einstein's work allowed a class of theories with a quadratic torsion in Lagrangian density \cite{MW77,M78,HS79}. Other trials to obtain a gauge field theory of gravity using the teleparallel geometry have shown a great interest \cite{U56,K61,S64,H76,U80}. Recent developments attempted a global approach by using arbitrary moving frames instead of the local expressions in the natural basis \cite{NS07,NS13}. Also, it worths to mention other developments to the teleparallel geometry by imposing Finslerian properties to this geometrical structure \cite{W09,WK11,YST12}. Actually, this geometry provides an alternative tool for deeper studies of the gravity.

An interesting variant of the TEGR is the Born-Infeld-modified teleparallel gravity. Within this framework the early cosmic acceleration (inflation) could be accounted for with no need to an inflaton field \cite{FF07,FF08}. Another remarkable variant on generalizations of TEGR are the $f(T)$ theories similar to the $f(R)$ extensions of Einstein-Hilbert action. Within this new class of modified gravity theories a particular form has been proposed to explain the late-time cosmic speeding up without dark energy (DE) \cite{BF09,L10,1008.4036,1011.0508}. As is well known that $f(R)$ theories are conformally equivalent to Einstein-Hilbert action plus a scalar field. In contrast, the $f(T)$ theories cannot be conformally equivalent to TEGR plus a scalar field \cite{Y2011}. Short period these pioneering studies have been followed by a large number of papers exploring different aspects of the $f(T)$ gravity in astrophysics \cite{CDDS11,FF011,FF11,IS12,CGS13,Nashed1,Nashed2,RHTMM13,Nashed3,BFG15,Nashed4,Nashed5} and in cosmology \cite{1205.3421,BNO14,BO14,JMM14,HLOS14,NH14,WH14,HN14,1503.05281,1503.07427}. Some applications show interesting results, e.g. avoiding the big bang singularity by presenting a bouncing solution \cite{CCDDS11,CQSW14}. Also, a graceful exit inflationary model within the $f(T)$ cosmology has been argued in \cite{NHS14}. A recent promising variant is the modified teleparallel equivalent of Gauss-Bonnet gravity and its applications \cite{KS114,KS214,KS314}.

Spatially flat universe (SFU), i.e. $k=0$, is used widely in the literature seeking for consistency with cosmic observations. However, recent observations by Planck satellite suggest, also, an open universe as a reliable model \cite{Pl2}. On the other hand, the smallness of the curvature density parameter cannot be covered by SFU assumption. In the present study, instead of restrict ourselves to SFU, we impose FLU assumptions into the modified $f(T)$ Friedmann equations. This model identifies a hidden class of $f(T)$ gravity theories that cannot be covered by assuming a SFU. The rest of this paper is structured as follows: In Section \ref{Sec2}, we briefly review the teleparallel geometry and the $f(T)$ gravity theories. In Section \ref{Sec3}, we use the modified version of the Friedmann equations according to $f(T)$ gravity to describe a FLU. In Section \ref{Sec4}, we discuss the physical and cosmological consequences of the obtained results. In Section \ref{Sec5}, we assume the case when the torsion tensor is made of a scalar field. This enables to construct an exact inflation model powered by a quintessence like field sensitive to the spacetime symmetry. In addition, it enables to induce a generalized Starobinsky potential from the obtained $f(T)$ gravity theory.  In Section \ref{Sec6}, we study possible solutions of the slow-roll parameters of this theory. We discuss testable predictions of the theory according to the recent released Planck and BICEP2 data \cite{Pl1, Pl2, BICEP2}. The work is summarized in Section \ref{Sec7}.
\section{Extended Teleparallel Gravity}\label{Sec2}
In what follows, we give the structure of Weitzenb\"{o}ck $4$-space. It is described as a pair $(M,~h_{a})$, where $M$ is a $4$-dimensional smooth manifold and $h_{a}$ ($a=0,\cdots, 3$) are $4$-linearly independent vector (vierbein) fields defined globally on $M$. Consequently, the $|h|:=\det(h_{a}^{\mu})$ is nonzero. The vierbein fields and their dual coframes are orthonormal, i.e.
$h_{a}{^{\mu}}h^{a}_{\nu}=\delta^{\mu}_{\nu}$ and $h_{a}{^{\mu}}h^{b}_{\mu}=\delta^{b}_{a}$, where $\delta$ is the Kronecker tensor. The metric space can be constructed from the vierbein fields $g_{\mu\nu}:=\eta_{ab}h^{a}{_\mu}h^{b}{_\nu}$ where $\eta_{ab}=\textmd{diag}(1,-1,-1,-1)$ is the Minkowski metric for the tangent space, so the Riemannian geometry is recovered. It is, also, possible to construct Weitzenb\"{o}ck nonsymmetric connection \begin{equation}\label{W_connection}
\Gamma^{\alpha}{_{\mu\nu}}:=h^{a}{_{\mu}}\partial_{\nu}h_{a}{^{\alpha}}=-h_{a}{^{\alpha}}\partial_{\nu}h^{a}{_{\mu}},
\end{equation}
where $\partial_{\nu}=\frac{\partial}{\partial x^{\nu}}$. The Weitzenb\"{o}ck space is characterized by the vanishing of the vierbein's covariant derivative, i.e.
\begin{equation}\label{AP_condition}
\nabla^{(\Gamma)}_{\nu}h^{a}{_{\mu}}:=\partial_{\nu}
{h_a}^\mu+{\Gamma^\mu}_{\lambda \nu} {h_a}^\lambda\equiv 0,
\end{equation}
where the covariant derivative $\nabla^{(\Gamma)}$ is w.r.t. the Weitzenb\"{o}ck  connection. So this property identifies auto parallelism or absolute parallelism condition. As a matter of fact, the $\nabla^{(\Gamma)}$ operator is not covariant under local Lorentz transformations (LLT) $SO(3,1)$ allowing all LLT invariant geometrical quantities to rotate freely in every point of the space \cite{M2013}. In this sense, the symmetric metric (10 degrees of freedom) cannot predict exactly one set of vierbein fields; then the extra six degrees of freedom of the 16 vierbein fields need to be fixed in order to identify exactly one physical frame. The frames  $h_{a}$ are called the parallelization vector fields. It can be shown that the teleparallelism condition (\ref{AP_condition}) implies the metricity condition $\nabla^{(\overcirc{\Gamma})}_{\sigma}g_{\mu\nu}\equiv 0$. The Weitzenb\"{o}ck connection (\ref{W_connection}) is curvature free while it has torsion tensor. The torsion $T$ and the contortion $K$ tensor fields of type (1,2) in spacetime coordinates are
\begin{eqnarray}
T^\alpha{_{\mu\nu}}&:=&{\Gamma^\alpha}_{\nu\mu}-{\Gamma^\alpha}_{\mu\nu}={h_a}^\alpha\left(\partial_\mu{h^a}_\nu-\partial_\nu{h^a}_\mu\right),\\
K^{\mu \nu}{_\alpha}  & := &-\frac{1}{2}\left({T^{\mu \nu}}_\alpha-{T^{\nu\mu}}_\alpha-{T_\alpha}^{\mu \nu}\right). \label{q4}
\end{eqnarray}
In the teleparallel space one may define three Weitzenb\"{o}ck invariants: $I_{1}=T^{\alpha\mu\nu}T_{\alpha\mu\nu}$, $I_{2}=T^{\alpha\mu\nu}T_{\mu\alpha\nu}$ and $I_{3}=T^{\alpha}T_{\alpha}$, where $T^{\alpha}=T_{\nu}{^{\alpha\nu}}$. We next define the invariant $T=AI_{1}+BI_{2}+CI_{3}$, where $A$, $B$ and $C$ are arbitrary constants \cite{M2013}. For the values: $A=1/4$, $B=1/2$ and $C=-1$  the invariant $T$ is just the Ricci scalar $R^{(\overcirc{\Gamma})}$, up to a total derivative term as mentioned in Section \ref{Sec1}; then a teleparallel version of gravity equivalent to GR can be achieved. The teleparallel torsion scalar is given in the compact form
\begin{equation}\label{Tor_sc}
T := {T^\alpha}_{\mu \nu}{S_\alpha}^{\mu \nu},
\end{equation}
where the superpotential tensor ${S_\alpha}^{\mu \nu}$ is defined as
\begin{equation}\label{q5}
{S_\alpha}^{\mu \nu}:= \frac{1}{2}\left({K^{\mu\nu}}_\alpha+\delta^\mu_\alpha{T^{\beta\nu}}_\beta-\delta^\nu_\alpha{T^{\beta \mu}}_\beta\right),
\end{equation}
which is skew symmetric in the last two indices. Similar to the $f(R)$ theory one can take the action of $f(T)$ theory as
\begin{equation}\label{q7}
{\cal L}({h^a}_\mu, \Phi_A)=\int |h|\left[\frac{M_{\textmd{\tiny Pl}}^2}{2}f(T)+{\cal L}_{m}(\Phi_A)\right]~d^{4}x,
\end{equation}
where $M_{\textmd{\tiny Pl}}$ is the reduced Planck mass, which is related to the gravitational constant $G$ by $M_{\textmd{\tiny Pl}}=\sqrt{\hbar c/8\pi G}$. Assuming the units in which $G = c = \hbar = 1$, in the above equation $ |h|=\sqrt{-g}$ and  $\Phi_A$ are the matter fields. The variation of (\ref{q7}) w.r.t. the tetrad field ${h^a}_\mu$ requires the following field equations \cite{BF09}
\begin{eqnarray}\label{q8}
\nonumber &&{S_\mu}^{\rho \nu} \partial_{\rho} T f_{TT}+\left[h^{-1}{h^a}_\mu\partial_\rho\left(h{h_a}^\alpha
{S_\alpha}^{\rho \nu}\right)-{T^\alpha}_{\lambda \mu}{S_\alpha}^{\nu \lambda}\right]f_T\\
&&-\frac{1}{4}\delta^\nu_\mu f=-4\pi{{\Theta}_{\mu}}^{\nu},
\end{eqnarray}
where $f \equiv f(T)$, $f_{T}=\frac{\partial f(T)}{\partial T}$, $f_{TT}=\frac{\partial^2 f(T)}{\partial T^2}$ and ${{\Theta}_{\mu}}^{\nu}$ is the energy-momentum tensor.
\section{Cosmological Modifications of $f(T)$}\label{Sec3}
We first assume that our universe is an isotropic and homogeneous, which directly gives rise to the tetrad field given by Robertson \cite{R32}. This can be written in spherical polar coordinate ($t$, $r$, $\theta$, $\phi$) as follows:
\begin{small}
\begin{eqnarray}\label{tetrad}
\nonumber \left({h_{a}}^{\mu}\right)=
\left(
  \begin{array}{cccc}
    1 & 0 & 0 & 0 \\
    0&\frac{L_1 \sin{\theta} \cos{\phi}}{4a(t)} & \frac{L_2 \cos{\theta} \cos{\phi}-4r\sqrt{k}\sin{\phi}}{4 r a(t)} & -\frac{L_2 \sin{\phi}+4 r \sqrt{k} \cos{\theta} \cos{\phi}}{4 r a(t)\sin{\theta}} \\[5pt]
    0&\frac{L_1 \sin{\theta} \sin{\phi}}{4 a(t)} & \frac{L_2 \cos{\theta} \sin{\phi}+4 r \sqrt{k}\cos{\phi}}{4 r a(t)} & \frac{L_2 \cos{\phi}-4 r \sqrt{k} \cos{\theta} \sin{\phi}}{4 r a(t)\sin{\theta}} \\[5pt]
    0&\frac{L_1 \cos{\theta}}{4 a(t)} & \frac{-L_2 \sin{\theta}}{4 r a(t)} & \frac{\sqrt{k}}{a(t)} \\[5pt]
  \end{array}
\right),\\
\end{eqnarray}
\end{small}
where $a(t)$ is the scale factor, $L_1=4+k r^{2}$ and $L_2=4-k r^{2}$. The tetrad (\ref{tetrad}) has the same metric as FRW metric
$$ds^2=dt^{2}-a^{2}(t) \left[\frac{dr^{2}}{1-\frac{1}{4}k r^{2}}+r^{2} d\theta^2+r^{2}\sin^{2}(\theta) d\phi^{2}\right].$$
\subsection{Modified Friedmann equations}\label{Sec3.1}
Applying the $f(T)$ field equations (\ref{q8}) to the FRW universe (\ref{tetrad}), assuming an isotropic perfect fluid  the energy-momentum tensor takes the form ${\Theta_{\mu}}^{\nu}=\textmd{diag}(\rho,-p,-p,-p)$. The $f(T)$ field equations (\ref{q8}) read
\begin{eqnarray}
\rho=\Theta_{0}^{~0}&=&\frac{1}{16 \pi}(f+12 H^2 f_{T}).\label{dens1}
\end{eqnarray}
\begin{eqnarray}
\nonumber p=\Theta_{(\alpha)}{^{(\alpha)}}&=&-\frac{1}{16 \pi}\left[(f+12 H^2 f_{T}) + 4\dot{H}(f_{T}-12 H^2 f_{TT})\right.\\
&-&\left.\frac{4k}{a^2}(f_{T}+12H^2 f_{TT})\right], \label{press1}
\end{eqnarray}
where $(\alpha)$ identifies the spatial coordinate running from $1$, $...$, $3$ and $H(:={\dot{a}}/{a})$ is the Hubble parameter, the dot denotes the derivative w.r.t. cosmic time $t$. Equations (\ref{dens1}) and (\ref{press1}) are the modified Friedmann equations in the $f(T)$ gravity. Substituting from the vierbein (\ref{tetrad}) into (\ref{Tor_sc}), we get the torsion scalar
\begin{eqnarray}\label{Tscalar}
\nonumber T&=&\frac{6 k- 6 \dot{a}^2}{a^2}\\
 &=&-6H^{2}\left(1+\Omega_{k}\right),
\end{eqnarray}
where $\Omega_{k}:=-\frac{k}{a^{2}H^{2}}$ is called the curvature density parameter. In above, the modified $f(T)$ Friedmann equations might show interesting results since the first round  bracket in (\ref{press1}) is the additive inverse of (\ref{dens1}). This provides a good chance to hunt a cosmological constant like matter perfectly, if the extra terms of (\ref{press1}) vanish. However, these terms enable evolutions away from the cosmological constant. By careful look to (\ref{press1}) we find that these extra quantities can be split into a free $k$ term and $k$ dependent term\footnote{One should note that $f(T)$ may contain $k$ dependent quantities.}. Taking the SFU assumption leads to disappearance of the last term in (\ref{press1}); then the chance for hunting cosmological constant like  DE is by taking the highly restrictive condition $\dot{H}=0$ or by taking its coefficient $f_{T}-12H^2f_{TT}=0$. Since we are dealing within the SFU framework, hence the teleparallel torsion scalar (\ref{Tscalar}) can be related to Hubble parameter by $T=-6H^{2}$. Only the SFU identifies a particular class of $f(T)\propto \sqrt{T}$. In this article we take a different path allowing evolution away from the cosmological constant without assuming spatial flatness, but enforcing the evolution to be flat like model.
\subsection{Searching for a flat like universe}\label{Sec3.2}
Most cosmological models assume SFU models seeking for a perfect match with cosmological observations. As a matter of fact, taking the spatial flatness as a firm prediction of inflation is not quite accurate. It is easy to show that the smallness of $\Omega_{k}$ can be a reflection of $k\ll a^{2}H^{2}$ as expected at early universe. It has been shown that even with a relatively large curvature parameter one can gain all advantages of inflation. On the other hand, it has been shown that the curvature density parameter average is $|\Omega_{k}|\lesssim 0.15$ at $1\sigma$ confidence \cite{FMR15}. Moreover, it has been shown that the so-called modified growth parameters are correlated with the curvature density parameter with a recognizable deviation at $|\Omega_{k}|\geq 0.05$, which leads some to conclude that the spatial curvature must be included in the analysis with other cosmological parameters \cite{DI12,NH14,ZZCZ14,HN14}. Recalling the $f(T)$ modified Friedmann equations (\ref{dens1}) and (\ref{press1}), we allow the fluid to evolve away from the cosmological constant like fluid regardless the value of $k$. Instead of taking the constraint of the SFU model, we assume the vanishing of the coefficient of $k$ in (\ref{press1}) so that
\begin{equation}\label{diff_F}
    a^2 f_{T}+12 \dot{a}^2 f_{TT}=0.
\end{equation}
The FLU model ansatz (\ref{diff_F}) identifies a hidden special class of $f(T)$ gravity theories which cannot be covered by taking SFU model. The solution of (\ref{diff_F}) is not easy in non-flat models. Also, it worths to mention that this treatment cannot be applied in the TEGR theory, i.e. $f(T)=T$, since the coefficient of  $k$ in (\ref{press1}) is a constant. So it would not provide us with a condition similar to (\ref{diff_F}) for examining a FLU. In this case one is obliged to assume a global spatial flatness by putting $k=0$ to study flat universe model. Thus we expect that the FLU treatment enables to study gravity beyond the TEGR or GR domains. As $f(T)$ in FRW spacetime is a function of time $f(T \rightarrow t)$, one easily can show that
\begin{equation}
  f_{T} = \dot{f}/\dot{T},~~  f_{TT} = \left(\dot{T} \ddot{f}-\ddot{T} \dot{f}\right)/\dot{T}^3.\label{Fd2T}
\end{equation}
Substituting from (\ref{Fd2T}) into (\ref{diff_F}), then solve  to $f(T \rightarrow t)$ we get:
\begin{equation}\label{fT1}
f(t)=\Lambda+\lambda\int {e^{{\int} {\frac{k^2+(3\ddot{a}a-5\dot{a}^2)k+2\ddot{a}^2a^2+4\dot{a}^4-7\dot{a}^2\ddot{a}a+\dot{a}\dddot{a}a^2}
{\dot{a}a(\ddot{a}a-\dot{a}^2+k)}} dt}}dt,
\end{equation}
where $\Lambda$ and $\lambda$ are integration constants. We later show that the constant $\Lambda$ can be perfectly interpreted as cosmological constant. In order to reduce the dependence on $k$, we follow the same ansatz (\ref{diff_F}) by taking a vanishing value of the coefficient of $k$. So, in the above equation, we take
\begin{equation}\label{assump2}
    3\ddot{a} a - 5 \dot{a}^2=0,
\end{equation}
by solving for $a(t)$ we get the scale factor
\begin{equation}\label{scale_factor}
    a(t)=a_{0}\left[\frac{3}{3-2H_{0}(t-t_{0})}\right]^{3/2},
\end{equation}
where $a_{0}$ is an arbitrary constant of integration, with an initial condition $H_{0}=H(t_{0})$. One should mention that the scale factor is independent of the choice of $k$. On the other hand, we do not expect or accept that the three world models $k=0,\pm 1$ to be completely coincide, then the function $f(T \rightarrow t)$ should depend on the choice of $k$ in its final form. Substituting from (\ref{scale_factor}) into (\ref{Tscalar}) we get the torsion scalar
\begin{equation}\label{Tsc}
    T(t)=\frac{6 k [3-2H_{0}(t-t_{0})]^{5}-1458 a_{0}^{2}H_{0}^{2}}{27 a_{0}^{2} [3-2H_{0}(t-t_{0})]^{2}}.
\end{equation}
Substituting from (\ref{scale_factor}) into (\ref{fT1}) and expanding the exponential around $t=0$; then the $f(T)$ is obtained as a function of $t$ as
\begin{equation}\label{f(t)}
    f(t)=\sum_{n=0}^{\infty}c_{n}t^{n},
\end{equation}
where $c_{n}$ are constant coefficients consist of the set of constants $\{k, a_{0}, t_{0}, H_{0}, \lambda, \Lambda\}$. The zeroth term of (\ref{f(t)}) derives the Friedmann equations (\ref{dens1}) and (\ref{press1}) to produce the cosmological constant DE, so fixes the value of the coefficient $c_{0}$ to the cosmological constant. We give examples of the $c_{n}$ coefficients: $c_0 = \Lambda$, $c_1=-2\tau_0 H_0 \lambda k - \frac{81a_0^2 \lambda}{8\tau_0^4 H_0^2}$, $c_2=-\frac{81a_0^2\lambda}{4\tau_0^5 H_0^2}-\frac{16\tau_0^5 H_0^4 \lambda k^2}{81a_0^2}$, ... etc, where $\tau_0=t_0+\frac{3}{2H_0}$. The higher orders of the expansion in (\ref{f(t)}) produce the evolution away from the cosmological constant. Equation (\ref{Tsc}) enables to write the time mathematically in terms of the torsion scalar, i.e. $t=\tau_0+\frac{3\sqrt{6}/2}{\sqrt{-T}}+O(\frac{1}{T^{3}})$; then we reexpress (\ref{f(t)}) in terms of the torsion scalar as inverse power law of $T$ as\footnote{We give the $f(T)$ in terms of the torsion scalar to explore the $f(T)$ in its usual form. But all the calculations in this work are performed using the time dependent form (\ref{f(t)}).}
\begin{equation}
\nonumber    f(T)\propto\sum_{n=0}^{\infty}\frac{\alpha_{n}}{\sqrt{-T^{n}}},
\end{equation}
where $\alpha_{n}$ are known constant coefficients consist of the same set of constants similar to the coefficients $c_{n}$. The above expression shows that $f(T)$ does not reduce to TEGR as expected so that it describes a non ordinary matter. Accordingly, we cannot assume a fixed EoS as in the classical cases. But it is more convenient to consider the case of the time dependent EoS corresponds to the contribution of the dynamical evolution of the $f(T)$ to the density and pressure. We summarize the FLU model as below:
\begin{itemize}
\item [(a)] In the GR theory, a fixed value of the equation of state (EoS) parameter, $\omega:=p(a(t))/\rho(a(t))$, is entered to the Friedmann equations as an input; then the scale factor $a(t)$ is obtained as an output.
\item [(b)] In the $f(T)$ theories, a fixed EoS parameter in addition to a scale factor $a(t)$ are entered to the Friedmann equations as inputs; then an $f(T)$ is obtained as an output.
\item [(c)] In this work which is governed by the $f(T)$ framework, we introduce two conditions (the FLU model assumptions, (\ref{fT1}) and (\ref{assump2})) as inputs; then we get a scale factor $a(t)$, $f(T)$ and a dynamical EoS parameter $\omega(t)$ as outputs.
\end{itemize}
In the (c) case,  the FLU constrains the scale factor $a(t)$ and $f(T)$ by the model assumptions.  These assumptions allow density and pressure to reformulate accordingly as functions of time. So it is not convenient to add an extra condition, e.g.  assuming a fixed EoS. Whereas the EoS in the FLU model should be treated as a time dependent output parameter, i.e. $\omega(t)=p(t)/\rho(t)$, this grantees a consistent system of Friedmann equations.  Actually, the only thing that we have to worry about, in $f(T)$ of cosmological applications, is obtaining compatible $a(t)$ and $f(T)$ \cite{NH14}. It worths to mention that the fine tuning problem facing DE models with a constant equation of state can be alleviated if we assume that the EoS is time dependent, e.g. quintessence models \cite{S02}. So one should think of this model as quintessence models rather than classical cosmological models.
\section{Cosmic Evolution}\label{Sec4}
In this Section, we perform a cosmological study of the FLU model to examine its capability to obtain the cosmic evolution. In addition, we study its consistency with the Friedmann equations. The FRW spacetime that is governed by $f(T)$ gravity can be determined by a scale factor $a(t)$ and an $f(T)$ form. As we mentioned before that the scale factor (\ref{scale_factor}) is independent of $k$, while the $f(t)$, namely (\ref{f(t)}), depends on $k$ as it should be. This allows to study different evolution scenarios according to the choice of $k$, assuming that the FRW universe is filled with a single fluid component given by (\ref{dens1}) and (\ref{press1}), we examine the cosmic evolution as follows.
\subsection{The large Hubble-spacetime}\label{Sec4.1}
We study the large $H$-regime, i.e. at early time universe. The Hubble parameter corresponds to the scale factor (\ref{scale_factor}) is given by
\begin{equation}\label{Hubble}
    H=\frac{3H_{0}}{3-2H_{0}(t-t_{0})},
\end{equation}
where the modified Friedmann equations are symmetric under $a \rightarrow -a$. We easily find that the universe is always accelerating as the deceleration parameter $q:=-\frac{a\ddot{a}}{\dot{a}^{2}}=-5/3$. Nevertheless, the matter density (\ref{dens1}) of this theory in the large $H$-spacetime is given as $\rho \rightarrow \frac{1}{16 \pi}\left(\Lambda+\frac{81 a_{0}^{2}H_{0}^{2}\lambda}{(3+2H_{0} t_{0})^{3}}\right)$, which agrees with the predictions of the vacuum density with a correction term. This is consistent with the inflationary universe scenario at this stage.
\subsection{Single-fluid equation of state}\label{Sec4.2}
In this $f(T)$ theory, we assume only a single fluid component to describe the matter of the universe. As we mentioned in Subsection \ref{Sec3.2}, there is a good chance to hunt negative pressure matter in the $f(T)$ framework with expected deviation from the cosmological constant behaviour. We now examine the nature of this single-fluid component and its possible evolution. Using (\ref{dens1}) and (\ref{press1}), we get a time dependent EoS parameter ($\omega := p/\rho$). The dynamical behaviour shows a similar asymptotic behaviour regardless the universe is flat or not. It is only affected by the order of expansion of (\ref{f(t)}). We need to mention that the $f_{n}(t)=c_{0}+c_{1}t+c_{2}t^{2}+ ...+c_{n}t^{n}$ shows that the leading term $f_{0}(t)=c_{0}=\Lambda$ producing an EoS as a constant function of time $\omega(t)=-1$, which describes the cosmological constant DE perfectly. However, the value of $\Lambda$ does not necessarily large as we will see later. We choose initial conditions to fit with the early universe, by taking a tiny initial scale factor combined with a large initial Hubble constant at Planck time $t_{0}=t_{\textmd{\tiny Pl}}$. When the first order correction is taken into consideration, the EoS approaches $\omega\rightarrow -7/9$ as $t \rightarrow t_{f}$, where $t_{f}\gg t_{\textmd{\tiny Pl}}$ is time large enough at which the EoS gets its final steady phase. While the second order correction allows the EoS to evolve up to $\omega \rightarrow -5/9$ as $t\rightarrow t_{f}$. We conclude that the series of the $f(t)$ up to the second order correction does not allow the EoS to crossover $\omega=0$ at anytime. These cases of $n=0,1,2$ are useful to describe the very early universe.

\begin{figure}
\begin{center}
\includegraphics[scale=.3]{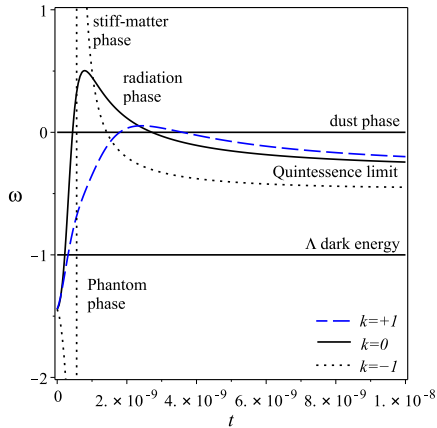}
\caption{The EoS up to $O(t^{3})$ evolution: The solid line represents the evolution in the flat space, while the dot  and dash lines represent the evolution in the open and closed models, respectively. The plots show almost the same initial phantom state and the same quintessence behaviour fate, which is in agreement with the expected evolution in the FLU model. The initial conditions are taken to fit with an early time phase as $a_{0}=10^{-9}$, $H_{0}=10^{9}~s^{-1}$, $t_{0}=t_{\textmd{\tiny Pl}}=10^{-44}~s$, $\Lambda=10^{-30}~s^{-2}$ and $\lambda=3$.}
\label{Fig1}
\end{center}
\end{figure}

We give the model $n=3$ more attention as it produces an excellent agreement with the acceptable cosmic evolution scenario, see the plots of Figure \ref{Fig1}. Although, we used a single-fluid component, its dynamical evolution allows the EoS parameter initially to start from  $\omega < -1$ (phantom). The evolution of the EoS shows almost the same behaviour in the three models. Then it evolves to pressureless cold dark matter (CDM) ($\omega=0$), radiation ($\omega=1/3$) and possibly stiff-matter ($\omega=1$), while it shows a quintessence fate  ${\omega} \rightarrow -1/3$ as $t \rightarrow t_{f}$ in all models. This shows a unique origin of early and late cosmic accelerating expansion. So we find that the third order correction is the most physical scenario. In the cases of $n \geq 4$, we have a physical motivation to study the series  up to $n=9$ only as the EoS of higher order produces $\omega >1$ asymptotically which represents unknown matter so far.

We find that the EoS asymptotic behaviour increases by a quantized value $\frac{2}{9}$ as $\left\{-1\right.$, $-\frac{7}{9}$, $-\frac{5}{9}$, $\left. ...\right\}$ as the $n$-th order increases by unity in the series (\ref{f(t)}).
It worths to mention that the FLU model shares the quintessence models an important result, a dynamical EoS. In addition, the theory predicts similar evolutions at early time in the three models $k=0,\pm 1$. Moreover, it forces the flat/non-flat models to have identical behaviours at a later time. Thus, the FLU model succeeds to describe an effectively flat solution without imposing a global vanishing sectional curvature by taking $k=0$.
\subsection{Conservative universe}\label{Sec4.3}
As mentioned in the introduction of the present Section that we extract a scale factor (\ref{scale_factor}) and an $f(T)$ (\ref{f(t)}) as solutions of two assumptions (\ref{fT1}) and (\ref{assump2}) of the FLU model, not directly from Friedmann equations as usual. We examine the validity of the Friedmann equations for this model. Actually, the case of the cosmological constant with a fixed EoS parameter ($\omega=-1$) that is required to describe the accelerating universe, one cannot get an evolutionary scenario for this dark component. In addition, its continuity equation
\begin{equation}\label{contn}
    \dot{\rho}+3H(\rho+p)=0,
\end{equation}
shows that $\dot{\rho}=0$, i.e. density is constant. Nevertheless, the density should decreases as universe expands. This conflict implies a continuous creation of matter in order to keep the density of the expanding universe constant. This case is similar to the steady state cosmology which breaks the conservation principle. This problem vanishes as a consequence of assuming a dynamical EoS as in our case here of the FLU model. In this $f(T)$ theory, we use one fluid component to describe the evolution of the universe, with a negative pressure matter dominating most of different epochs during the early time evolution. Moreover, we get a conservative universe. This can be obtained by substituting from (\ref{scale_factor}) and (\ref{f(t)}) into (\ref{dens1}) and (\ref{press1}), which shows that the continuity equation is always verified up to any order of expansion of (\ref{f(t)}).

We summarize the results of Section \ref{Sec4} as follows: The FLU model (i) is consistent with inflationary scenario universe, (ii) has dynamical EoS parameter allows cosmic evolution, and (iii) satisfies the continuity equation so Friedmann. It is well known that the scalar field plays the key role in the quintessence models to interpret the inflation stage at early universe. On the other hand, the torsion plays the main role in the teleparallel geometry. We next consider the role of the torsion as added value quality of the spacetime when formed by a scalar field and its role at the early universe time.
\section{Torsion Potential}\label{Sec5}
In order to get the idea of the torsion potential, we first discuss the physical meaning of the connection coefficients as the displacement field. So we use (\ref{q4}) to reexpress the Weitzenb\"{o}ck connection (\ref{W_connection}) as
\begin{equation}\label{contortion}
    {\Gamma^\mu}_{\nu \rho }=\overcirc{\Gamma}{^{\mu}}{_{\nu \rho}}+K^{\mu}{_{\nu \rho}}.
\end{equation}
By careful look to the above expression of the new displacement field, it consists of two terms. The first is the Levi-Civita connection which consists of the gravitational potential (metric coefficients, $g_{\mu \nu}$) and its first derivatives w.r.t. the coordinates. Where the second term is the contortion which consists of the tetrad   field and its first derivatives w.r.t. the coordinates. In this sense we find the first term contributes to the displacement field as the usual attractive force of gravity, while the second term contributes as a force too. While equation of motion tells that this force is repulsive \cite{W2012}. Now we can see how teleparallel geometry adds a new quality (torsion or contortion) to the spacetime allowing repulsive side of gravity to showup.
\subsection{Torsion potential of a scalar field}\label{Sec5.1}
We consider here the physical approach to construct the torsion from a scalar field $\varphi(x)$. In the view of the above discussion, we can treat the contortion in the Weitzenb\"{o}ck connection (\ref{contortion}) as a force. Accordingly, it is required to construct the contortion (or torsion) from a tensor and it first-order derivatives. Then, this tensor now plays the role of the potential of the torsion. We follow the approach that has been proposed by \cite{XSH96}, by introducing sixteen fields $t^{\mu}{_{a}}$ that are called $``$\textit{torsion potential}$"$. These fields form a quadruplet basis vectors, so we write the following linear transformation:
\begin{equation}
\nonumber    h_{a}=t^{\mu}{_{a}}\partial_{\mu},~~h^{a}=t^{a}{_{\mu}}dx^{\mu},
\end{equation}
the torsion potential $t^{\mu}{_{a}}$ and its inverse are satisfying the conditions:
\begin{equation}
\nonumber    t=\textmd{det}(t^{\mu}{_{a}})\neq 0,~~t^{\mu}{_{a}}t^{a}{_{\nu}}=\delta^{\mu}_{\nu},~~t^{\mu}{_{a}}t^{b}{_{\mu}}=\delta^{b}_{a}.
\end{equation}
Finally, this enables to express the torsion as \cite{XSH96}
\begin{equation}\label{torsion_pot}
    T^{\alpha}{_{\mu \nu}}=t^{\alpha}{_{a}}\left(\partial_{\mu}t^{a}{_{\nu}}-\partial_{\nu}t^{a}{_{\mu}}\right).
\end{equation}
The torsion potential $t^{\mu}{_{a}}$ can be reformed by a physical scalar, vector or tensor fields. This may have a great interest in physical applications. Now we take the case when the torsion potential is formed by a scalar field $\varphi(x)$ by taking
\begin{equation}\label{torsion_pot2}
t^{a}{_{\mu}}=\delta^{a}_{\mu}e^{\sqrt{3/2}\varphi},~~t^{\mu}{_{a}}=\delta^{\mu}_{a}e^{-\sqrt{3/2}\varphi},
\end{equation}
where $\varphi$ is a non-vanishing scalar field. Then the torsion and the contortion (\ref{q4}) can be reexpressed respectively as
\begin{eqnarray}
    T^{\alpha}{_{\mu\nu}}&=&\sqrt{3/2}\left(\delta^{\alpha}_{\nu}\partial_\mu \varphi-\delta^{\alpha}_{\mu}\partial_\nu \varphi\right),\label{semi-symm-torsion}\\
    K^{\mu\nu}{_{\alpha}}&=&\sqrt{3/2}\left(\delta^{\nu}_{\alpha}\partial^\mu \varphi-\delta^{\mu}_{\alpha}\partial^\mu \varphi\right),\label{semi-symm-contortion}
\end{eqnarray}
where $\partial^\mu \varphi=g^{\mu \alpha} \partial_\alpha \varphi$.
The above expressions have been used in order to formulate a theory of the electromagnetism minimally coupled to torsion and satisfies gauge invariance and minimal coupling principles \cite{R74,HRR78,H90,HO01}. Using (\ref{q5}), (\ref{semi-symm-torsion}) and (\ref{semi-symm-contortion}), the teleparallel torsion scalar (\ref{Tor_sc}) can be written in terms of the scalar field $\varphi$ as
\begin{equation}\label{Tsc_phi}
    T=-9\partial_\mu\varphi~ \partial^\mu\varphi.
\end{equation}
The above treatment shows that the torsion acquires dynamical properties and it  propagates through space. It worths to mention that the energy momentum tensor of a free scalar field represents a source of the dynamical torsion in the teleparallel description of gravity \cite{AP98}, which is not in agreement with the common belief that the spin matter only can produce torsion \cite{H76}. One should mention here that same results can be achieved by combining the conformal transformation of the tetrad fields $h^{\mu}{_{a}}\rightarrow e^{\varphi}{h}^{\mu}{_{a}}$ and $h^{a}{_{\mu}}\rightarrow e^{-\varphi}{h}^{a}{_{\mu}}$ in addition to the so-called Einstein $\lambda$-transformation (projective transformation) of the connection coefficients \cite{E55, FRM13} as
\begin{equation}\label{lambda-trans}
    \Gamma^{\alpha}{_{\mu \nu}} \rightarrow {\Gamma}^{\alpha}{_{\mu \nu}}-\delta^{\alpha}_{\nu} \partial_\mu \varphi.
\end{equation}
Actually, this approach, indeed, has a geometrical framework as well. Where the connection is assumed to be a semi-symmetric one, this case has been studied by many authors, c.f. \cite{NA2005}.
\subsection{Gravitational quintessence model}\label{Sec5.2}
It is well known that the cosmic inflation is powered by a scalar field $\varphi$.  On the other hand, the cosmological applications of  $f(T)$ gravity  show strong evidences of a cosmic inflation.  So there must be a link between these two descriptions, we see that equation (\ref{Tsc_phi}) provides this link as the teleparallel torsion scalar appears as a gradient of a scalar field. So it acquires dynamical properties and propagates through space. Also, equation (\ref{Tsc_phi}) enables us to map the density and pressure contributions in the Friedmann equations into the scalar field ($\rho\rightarrow \rho_{\varphi}$, $p\rightarrow p_{\varphi}$). This leads to reformulate the Friedmann equations of the torsion contribution as an inflationary background in terms of the scalar field $\varphi$. As mentioned before the FLU model provides $f(T)\propto \sum \frac{1}{\sqrt{-T^{n}}}$ does not reduce to TEGR theory of the ordinary matter. This is in agreement with the inflationary epoch where the matter contribution can be negligible. Accordingly, we consider the Lagrangian density of a homogeneous (real) scalar field $\varphi$ in potential $V(\varphi)$
\begin{equation}\label{lag_dens}
    \mathcal{L}_{\varphi}=\frac{1}{2}\partial_\mu \varphi~ \partial^\mu \varphi-V(\varphi).
\end{equation}
where the first expression in (\ref{lag_dens}) represents the kinetic term of the scalar field, as usual, while $V(\varphi)$ represents the potential of the scalar field. The variation of the action w.r.t. the metric $g_{\mu \nu}$ enables to define the energy momentum tensor as
\begin{equation}
 \nonumber   \mathcal{T}^{\mu \nu} = \frac{1}{2}\partial^\mu \varphi~ \partial^\nu \varphi-g^{\mu \nu} \mathcal{L}_{\varphi}.
\end{equation}
The variation w.r.t. the scalar field reads the scalar field density and pressure respectively as
\begin{equation}\label{press_phi1}
    \rho_{\varphi}=\frac{1}{2}\dot{\varphi}^2+V(\varphi),\quad \quad p_{\varphi}=\frac{1}{2}\dot{\varphi}^2-V(\varphi).
\end{equation}
The Friedmann equation (\ref{dens1}) of the non-flat models in absence of matter becomes
\begin{equation}\label{Hubble_sc}
\nonumber    H^2=\frac{8\pi}{3}\left(\frac{1}{2}\dot{\varphi}^2+V(\varphi)\right)-\frac{k}{a^{2}}.
\end{equation}
Using (\ref{press_phi1}) it is an easy task to show that continuity equation (\ref{contn}) reduces to the Klein-Gordon equation of homogeneous scalar field in the expanding FRW universe
\begin{equation}
\nonumber    \ddot{\varphi}+3H\dot{\varphi}+V'(\varphi)=0,
\end{equation}
where the prime denotes the derivative w.r.t. the scalar field $\varphi$. In conclusion, equation (\ref{Tsc_phi}) enables to define a scalar field sensitive to the vierbein field, i.e. the spacetime symmetry. In addition, equation (\ref{press_phi1}) enables to evaluate an effective potential from the adopted $f(T)$ gravity theory. The mapping from the torsion contribution to scalar field fulfills the Friedmann and Klein-Gordon equations. So the treatment meets the requirements to reformulate the torsion contribution in terms of a scalar field without  attempting a conformal transformation.
\subsection{Generalized Starobinsky potential by $f(T)$}\label{Sec5.3}
In the following treatment, we take the simple case of the flat space universe $k=0$ in order to compare the obtained results to the standard treatment of the scalar field theory in cosmology. Using (\ref{Tsc}), (\ref{Tsc_phi}) we get
\begin{equation}\label{kin}
\dot{\varphi}^2=\frac{3/2}{(t-\tau_{0})^2},
\end{equation}
where $\tau_{0}$ is as given in Subsection \ref{Sec3.2}. Integrating the above equation we write the scalar field $\varphi$ as
\begin{equation}\label{phi}
    \varphi=\varphi_{0} \pm \sqrt{6}/2\ln{(t-\tau_{0})},
\end{equation}
and $\varphi_{0}=\varphi(t_{0})$ is a constant of integration. The above equation allows to express the time $t$ mathematically in terms of the scalar field $\varphi$. So all the dynamical expressions can be expressed in terms of the scalar field.

The relation between the scalar fields $T$ and $\varphi$ has to be investigated. Using (\ref{Tsc}) and (\ref{phi}) we can express the torsion scalar field $T$ in terms of the scalar field $\varphi$ as
\begin{equation}\label{canon}
    T(\varphi)=-\frac{27}{2} e^{\pm 2\sqrt{2/3}(\varphi-\varphi_{0})}.
\end{equation}
As is well known at the strong coupling condition the inflaton field is related to the canonical scalar field $\Omega$ by \cite{KLR214}
\begin{equation}\label{canonical}
\varphi=\pm \sqrt{\frac{3}{2}}\log{\Omega}.
\end{equation}
Comparing the above equation with the teleparallel torsion (\ref{canon}) we get the transformation $T=-\xi \Omega^{2}$ where the constant $\varphi_{0}$ can be absorbed in the coefficient $\xi$. This indicates that the teleparallel torsion scalar might play a role similar to the canonical scalar field normally used in scalar-tensor theories of gravitation. Also, using (\ref{scale_factor}) and (\ref{f(t)}), the pressure (\ref{press1}) can be reexpressed in terms of the scalar field $\varphi$ as
\begin{equation}\label{dens_phi}
p_{\varphi}=-\frac{\Lambda }{16 \pi}-\lambda \sum_{n=0} \beta_{n} e^{-{n\sqrt{2/3}}(\varphi-\varphi_{0})},
\end{equation}
where $\beta_{n}$ are known constant coefficients. Similarly, we can evaluate the scalar field density (\ref{press_phi1}). With simple calculations one can find that the continuity equation of the scalar field still valid. Substituting from (\ref{kin}) and (\ref{dens_phi}) into (\ref{press_phi1}), we evaluate the potential of the scalar field $\varphi$ as
\begin{equation}\label{sc_pot}
V(\varphi)=V_{0}+\lambda \sum_{n=0} \beta_{n} e^{-n\sqrt{2/3}(\varphi-\varphi_{0})},
\end{equation}
where $V_{0}=\frac{\Lambda }{16\pi}+\frac{3}{4} e^{2\sqrt{2/3}(\varphi-\varphi_{0})}$. The higher orders contribution in (\ref{sc_pot}) are due to the order of expansion of the $f(T)$ which allow different potential types of slow-roll inflation models. So we find that both kinetic and potential energies are formed from the teleparallel torsion and $f(T)$ gravity, respectively. It is well known that the scalar field potential of the cosmic inflation has many different forms according to the assumed model, e.g. polynomial
chaotic, power law, natural, intermediate, ... etc. In this work, we provide another approach to construct the potential from the $f(T)$ gravity theory. According to the order of expansion of the potential (\ref{sc_pot}), this gives either Starobinsky-like inflation model \cite{St80}, where the potential blows up at $\varphi < 0$ and inflation can occur only at $\varphi>0$, or it gives a quadratic-like inflation model where the potential allows inflation to occur at both $\varphi < 0$ and $\varphi>0$. This will be discussed in more details as below.
\subsection{Classified torsion scalar potentials}\label{Sec5.4}
In the previous Subsection we applied a new technique to induce the scalar potential (\ref{sc_pot}) by the $f(T)$ gravity (\ref{f(t)}). In this theory we obtained a power series potential of $e^{-\sqrt{2/3}\varphi}$ where $\varphi$ represents the inflaton field. So the theory may cover different classes of inflationary models. We next examine the potentials correspond to the order of the expansion of (\ref{f(t)}). In this way, and with the help of the evaluated EoS induced by the fluid (\ref{dens1}) and (\ref{press1}), we might be able to compare these potentials to the already known inflaton potentials. The \textit{triple} (\ref{f(t)}), EoS and (\ref{sc_pot}) would enable us to give a physical classification of these potentials.
\subsubsection{The $V_{0}$-model}
\begin{figure}
\centering
\subfigure[$~n=0$]{\label{fig2a}\includegraphics[scale=.19]{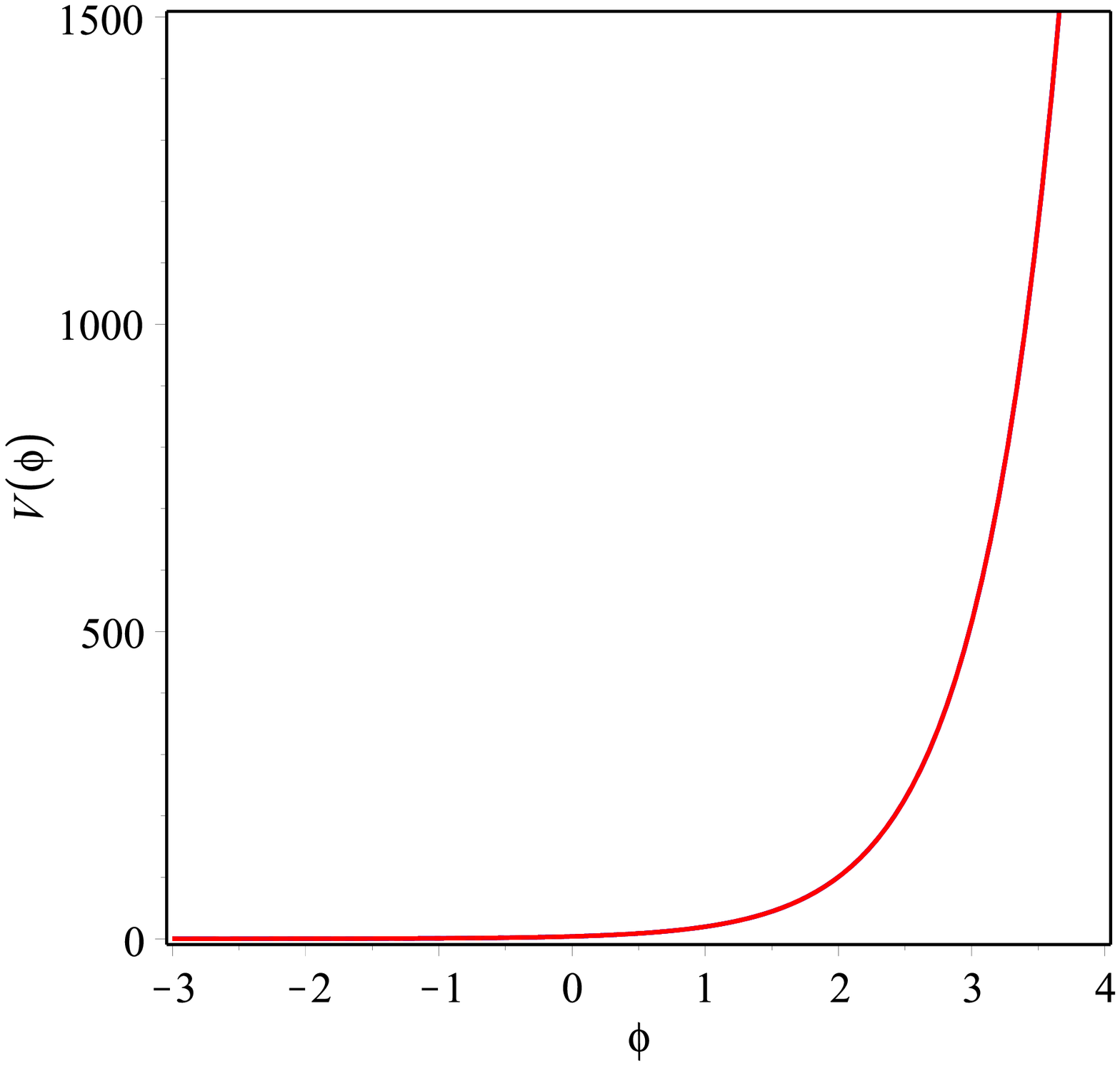}}
\subfigure[$~n=1$]{\label{fig2b}\includegraphics[scale=.19]{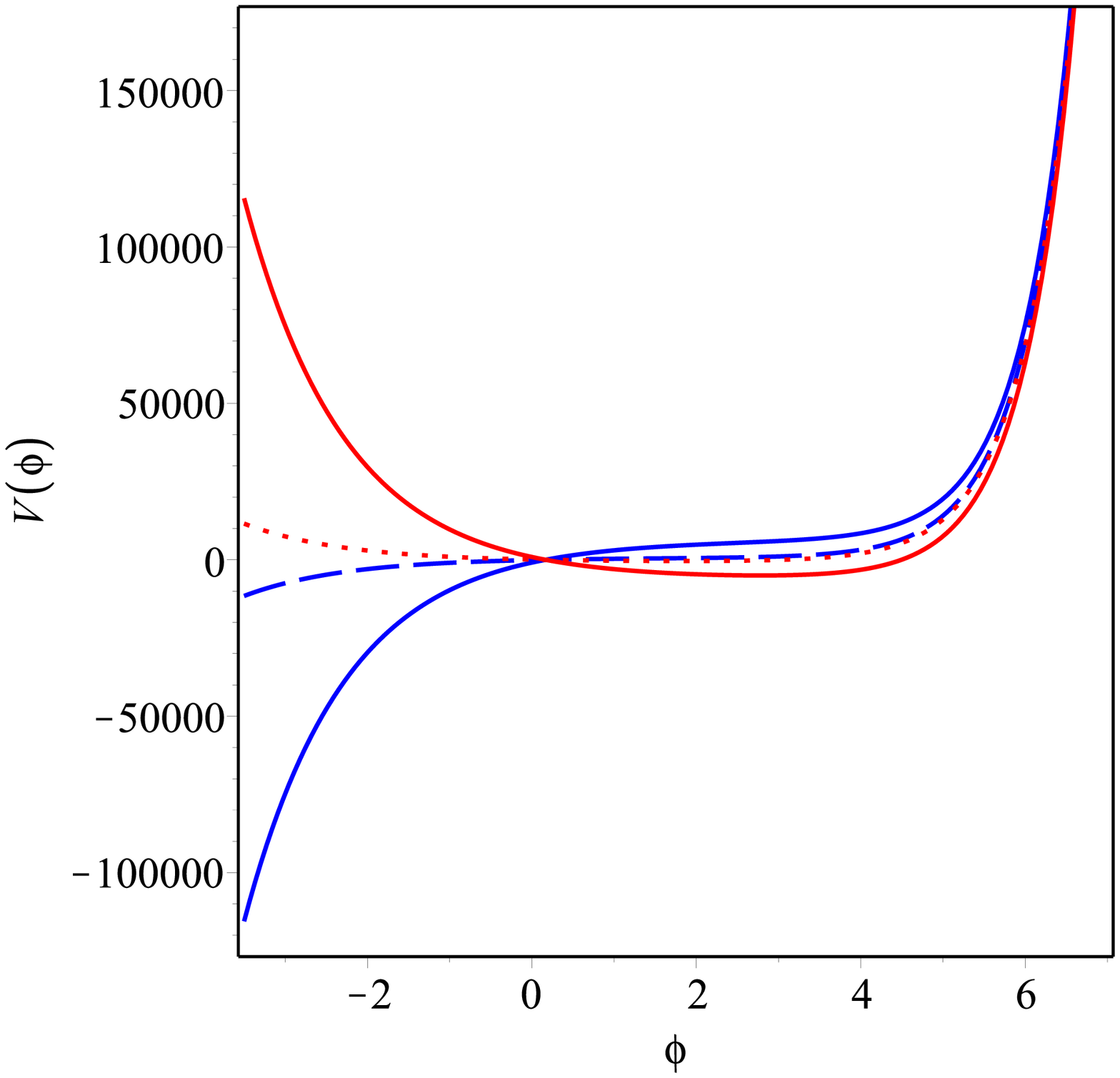}}
\subfigure[$~n=2$]{\label{fig2c}\includegraphics[scale=.19]{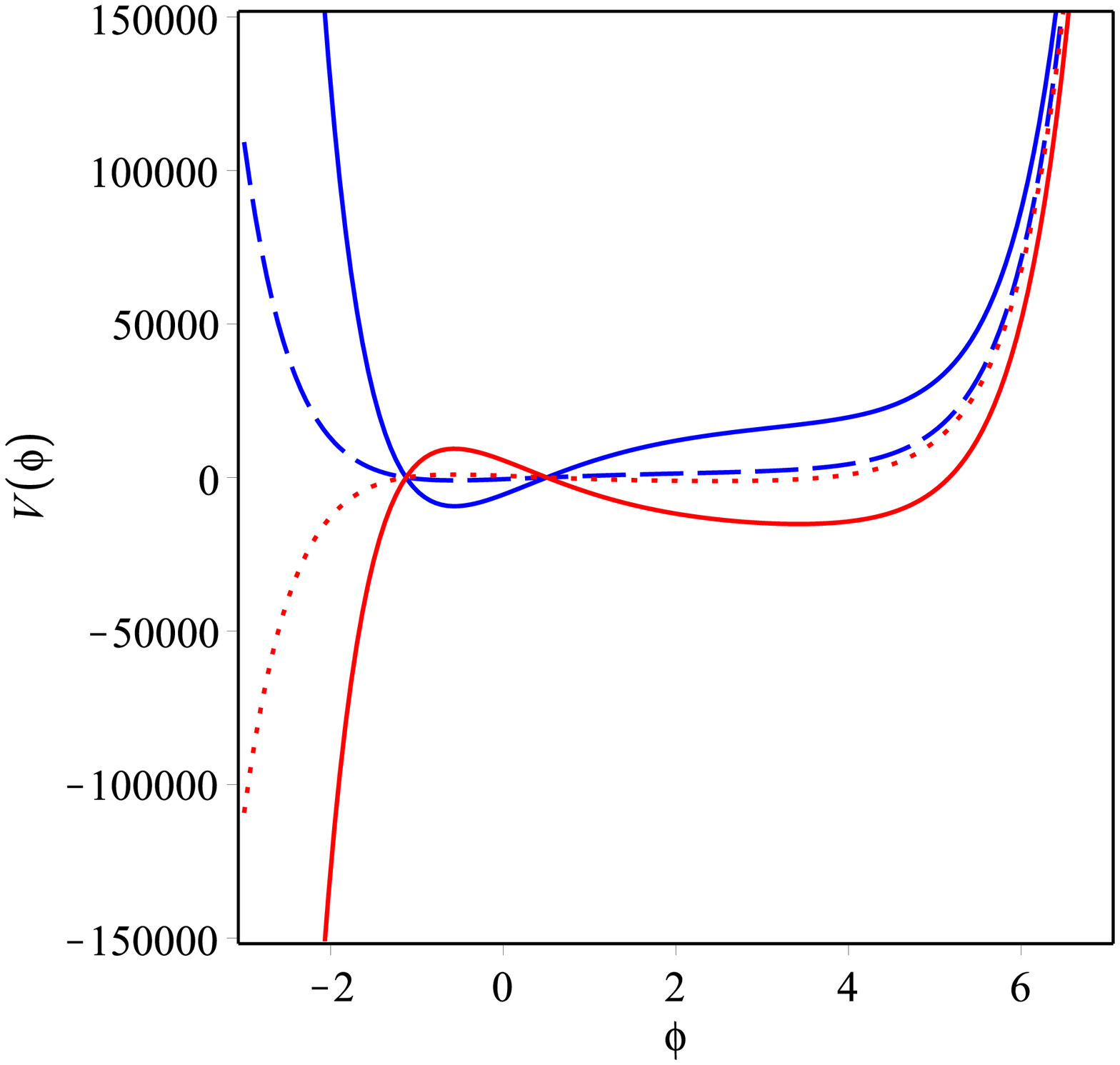}}
\subfigure[$~n=3$]{\label{fig2d}\includegraphics[scale=.19]{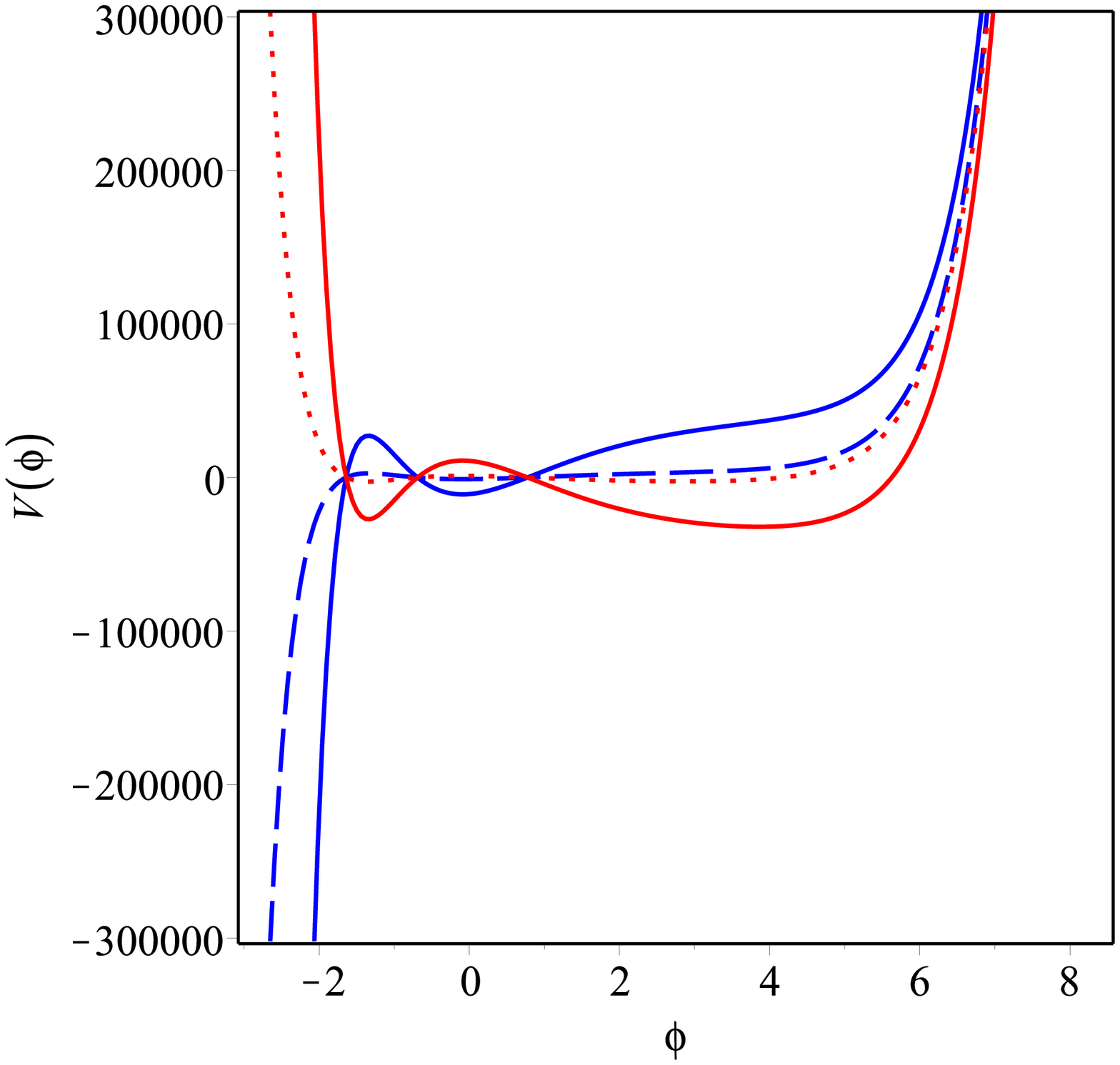}}
\subfigure[$~n=4$]{\label{fig2e}\includegraphics[scale=.19]{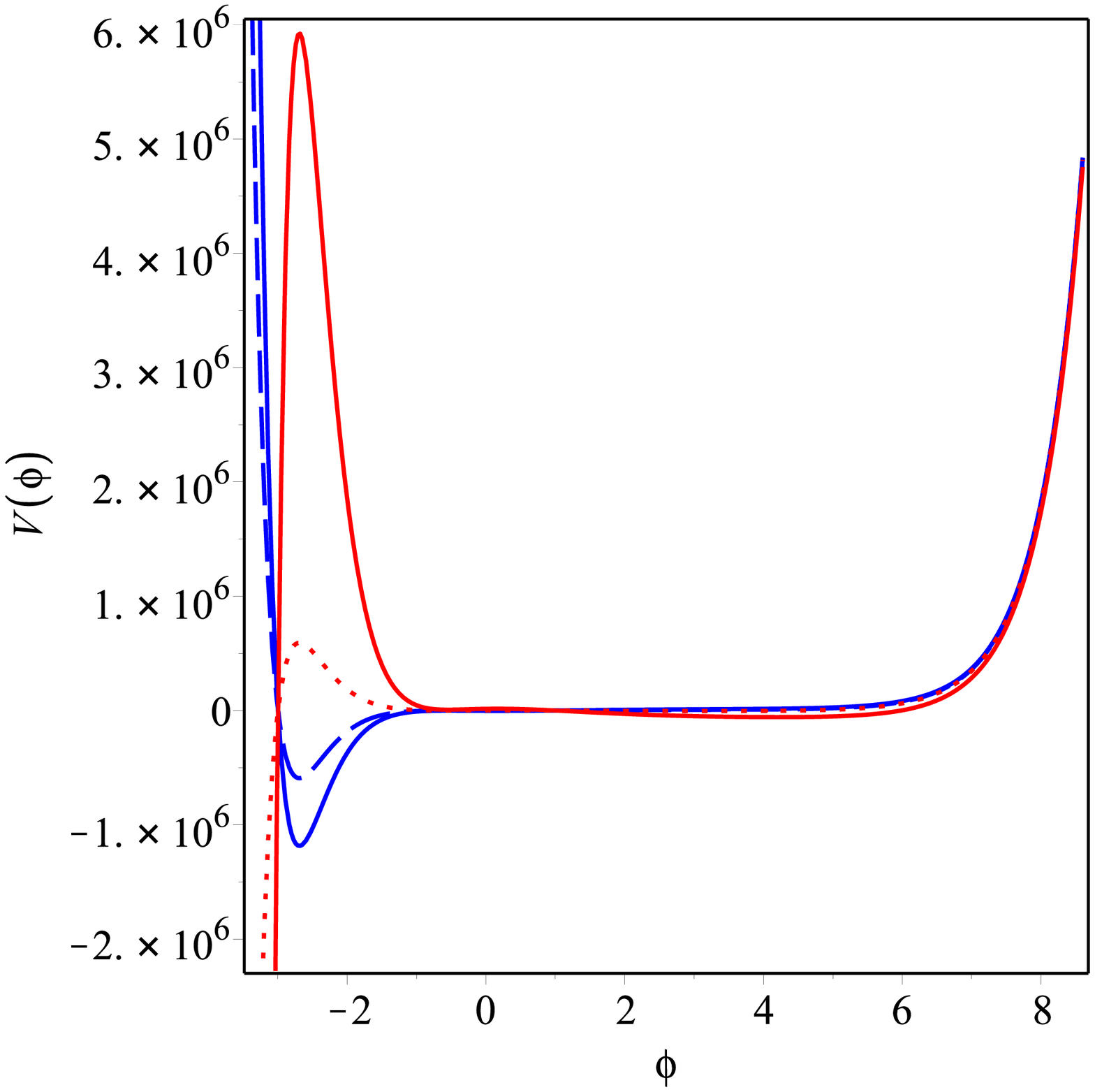}}
\subfigure[$~n=5$]{\label{fig2f}\includegraphics[scale=.19]{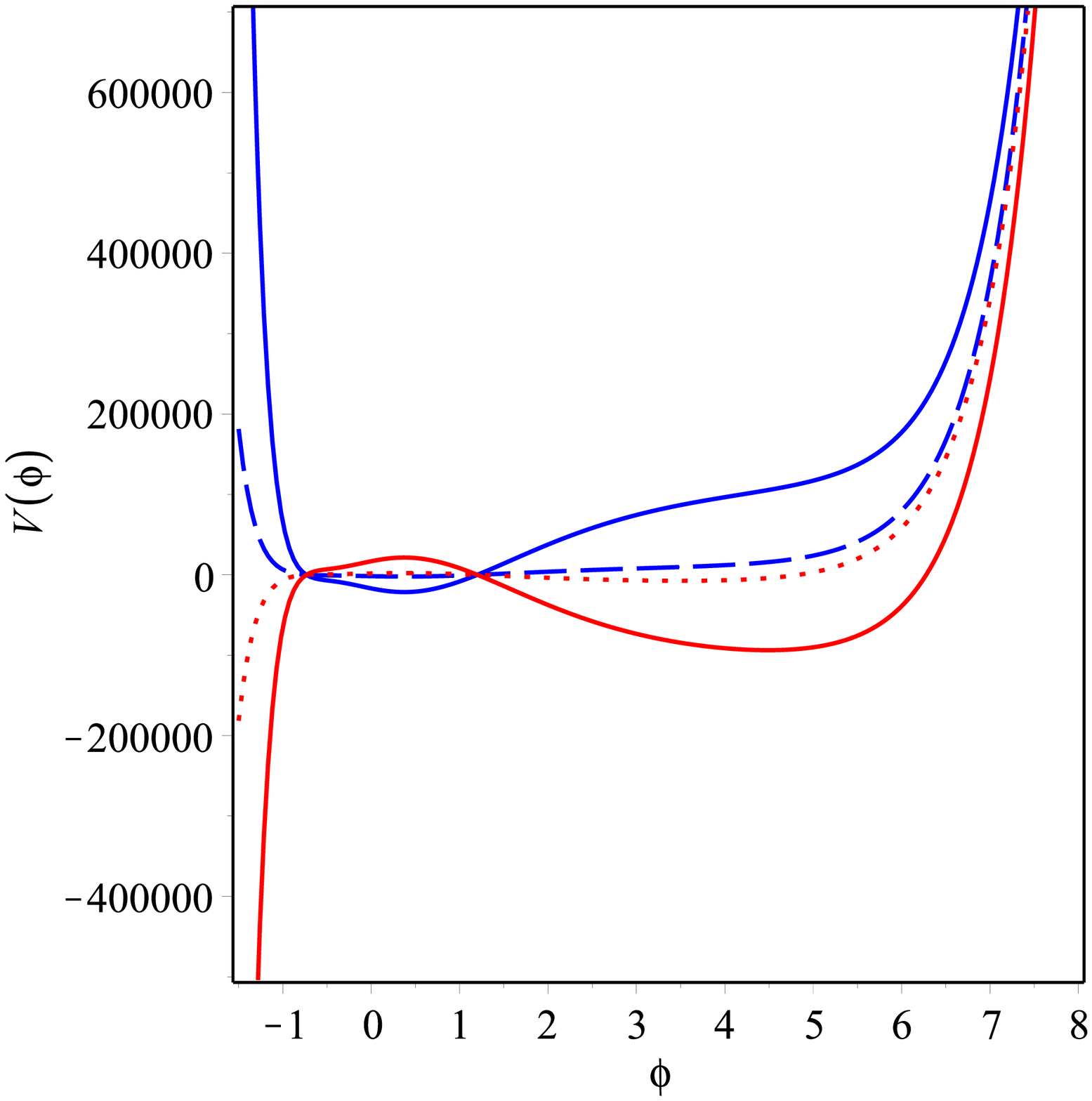}}
\subfigure[$~n=6$]{\label{fig2g}\includegraphics[scale=.19]{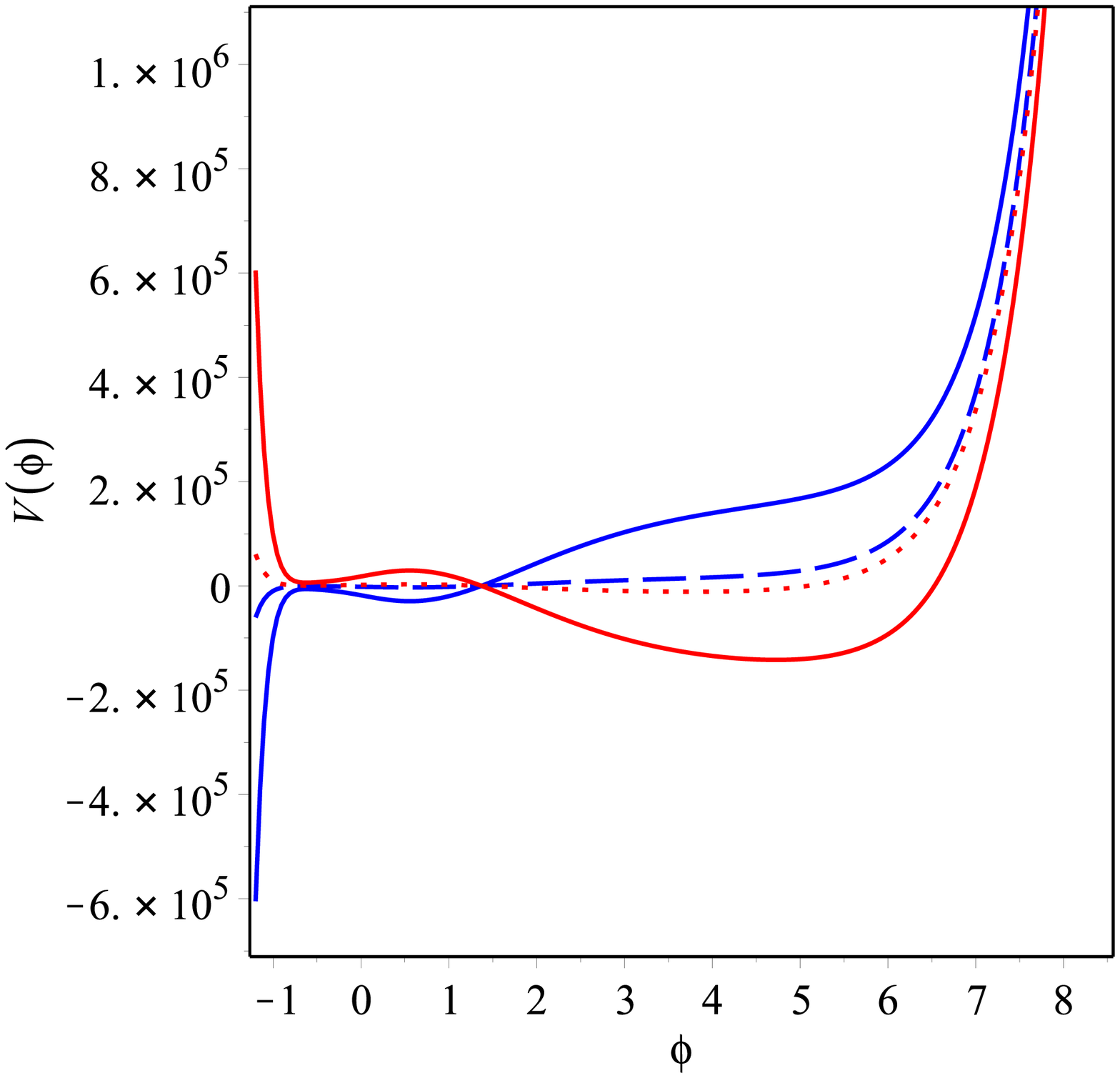}}
\subfigure[$~n=7$]{\label{fig2h}\includegraphics[scale=.19]{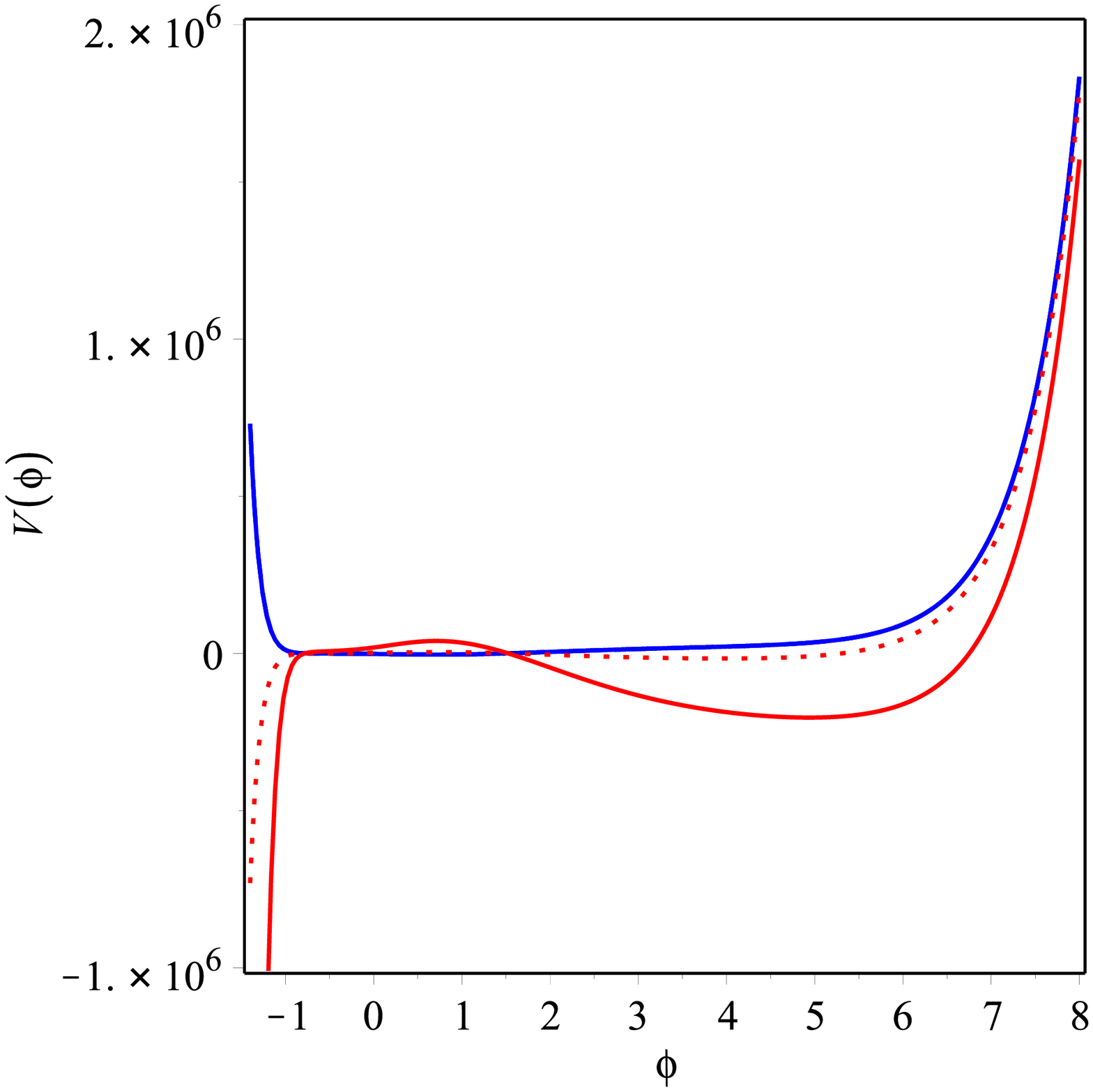}}
\caption[figtopcap]{Schematic plots of the scalar field potentials according to the order of expansion of the $f(T)$ function. The set of parameters $\{t_{0},a_{0},H_{0},\Lambda\}$ are taken as in Figure \ref{Fig1}. The blue solid lines are for $\lambda>1$, the blue dash lines are for $0<\lambda<1$, the red dot lines are for $0>\lambda>-1$ and the red solid lines are for $\lambda<-1$.}
\label{Fig2}
\end{figure}

The modified Friedmann equations (\ref{dens1}) and (\ref{press1}) of the $f(T)$ gravity theories pay attention to the choice of $f(T)=const.$ as it acts perfectly as the cosmological constant. This is exactly the case here, when assuming the zeroth order of (\ref{f(t)}) solution. We first evaluate the above mentioned \textit{triple} as
\begin{eqnarray}
    f_{0}(T)&=&\Lambda,\\
    \omega_{0}&=&-1,\\
    V_{0}&=&\frac{\Lambda }{16\pi}+\frac{3}{4} e^{2\sqrt{2/3}(\varphi-\varphi_{0})}, \label{V0}
\end{eqnarray}
where $-\infty < \varphi < \infty$. It is clear that $f_{0}(T)$ acts perfectly as cosmological constant DE ($\Lambda$DE) with EoS $\omega=-1$. A generic potential of this type can be obtained by adding a constant to exponential potential (power law inflation), see Figure \ref{Fig2}\subref{fig2a}. We will see later this potential pattern is recommended to perform $E$ and $B$ modes of the power spectrum polarization. The first term in (\ref{V0}) is the vacuum energy density (cosmological constant) of the false vacuum spacetime, while the second term associated to the vacuum potential represents a phase transition potential dragging the universe away from false vacuum ($\varphi=0$ state) to a true vacuum ($\varphi \neq 0$ state) at its minimum effective potential. However, the EoS still $\omega_{0}=-1$ during the whole stage. Now, we investigate the phase transition potential in (\ref{V0}). Using the relation (\ref{canonical}) we write the $V_{0}$ potential in terms the canonical scalar field as \[V_{0} \propto {1/\Omega^{2}},\] which gives an inverse square law potential. This type of potentials represents the original class of quintessence fields. Indeed, the later expression shows a minimum effective potential only at $\Omega>0$ or $\Omega < 0$ with a potential barrier at $\Omega=0$, while (\ref{V0}) allows the full range $-\infty < \varphi < \infty$. But as the inflation occurs only in a single plateau, then the two expression are identical at $\varphi > 0$.
\subsubsection{The $V_{1}$-model}
We take the expansions up to the first order giving the \textit{triple}
\begin{eqnarray}
    f_{1}(T)&=&\Lambda-\frac{a_{0}^{2}\lambda}{H_{0}^{2}\tau_{0}^{3}}\left[\frac{81}{8}
    +\frac{243\sqrt{6}}{16\tau_{0}\sqrt{-T}}\right],\\
    \lim_{t\rightarrow t_{f}}\omega_{1}&=&-7/9,\\
    V_{1}&=&V_{0}-\frac{a_{0}^{2}\lambda}{H_{0}^{2}\tau_{0}^{3}}\left[\frac{81}{128\pi}
    +\frac{63/\tau_{0}}{64\pi}e^{-\sqrt{2/3}(\varphi-\varphi_{0})}\right]. \label{V1}
\end{eqnarray}
The $f_{1}(T) \propto \frac{1}{\sqrt{-T}}$ which is usually used to identify the CDM. In this model, the additional terms of the potential $V_{1}$, given by (\ref{V1}), contribute to decrease the potential relative to the kinetic energy. This allows the EoS to evolve above the $\Lambda$DE value by $2/9$, i.e. $\omega_{1}\rightarrow -7/9$, which is in agreement with the previous results of Subsection \ref{Sec4.2}. Moreover, the potential $V_{1}$ reduces to $V_{0}$ where $\lambda=0$. However, the nonvanishing values of $\lambda$ show interesting patterns, there plots appear in Figure \ref{Fig2}\subref{fig2b}, they predict similar behaviours at $\varphi>0$ where the potentials are nearly flat at the false vacuum ($\varphi=0$) slowly rolls to its effective minimum at the true vacuum ($\varphi \neq 0$). At $\varphi < 0$, the potentials predict different behaviours according to the value of $\lambda$. At very small negative or positive values of $\lambda$ the potentials are nearly flat. The large positive $\lambda$ produces Starobinsky-like model where the potential blows up exponentially. Nevertheless, the large negative $\lambda$ turns the potential to a quadratic-like model.
\subsubsection{The $V_{2}$-model}
With higher order expansion we have these \textit{triple}
\begin{eqnarray}
    f_{2}(T)  &=&\Lambda -\frac{a_{0}^{2}\lambda}{H_{0}^{2}\tau_{0}^{3}}\left[\frac{243}{8}
    +\frac{1215\sqrt{6}}{16\tau_{0}\sqrt{-T}}-\frac{2187}{8\tau_{0}^{2} T}\right],\\
    \lim_{t\rightarrow t_{f}}\omega_{2}&=&-5/9,\\
\nonumber V_{2}&=&V_{0}-\frac{a_{0}^{2}\lambda}{H_{0}^{2}\tau_{0}^{3}}\left[\frac{243}{128\pi}
    +\frac{315}{64\pi}e^{-\sqrt{2/3}(\varphi-\varphi_{0})}\right.\\
    &&+\left.\frac{135/\tau_{0}}{64\pi}e^{-2\sqrt{2/3}(\varphi-\varphi_{0})}\right].
\end{eqnarray}
This case still in the quintessence range where $\omega_{2} \rightarrow -5/9$. The potential $V_{2}$ reduces to $V_{0}$ at the limit $\lambda \rightarrow 0$ where the kinetic dominates over the potential. On the other hand, $V_{2}$ reduces to the original Starobinsky potential where kinetic energy is negligible during the inflation. The Starobinsky pattern is clear in Figure \ref{Fig2}\subref{fig2c} in the case $\lambda>1$ where the potential is dominant. The non-vanishing values of $\lambda$ produce almost the same behaviour at $\varphi > 0$, while they produce different behaviours at $\varphi < 0$. In contrast to the $V_{1}$, the large negative $\lambda$ allows Starobinsky-like model to dominate the $\varphi < 0$ epoch while the large positive $\lambda$ turns the potential to a quadratic-like model, see Figure \ref{Fig2}\subref{fig2c}, at $\varphi < 0$. Its plot shows two different minima allowing inflation in both $|\varphi| >0$.
\subsubsection{The $V_{3}$-model}
We study one more case where the \textit{triple} are given by
\begin{eqnarray}
\nonumber f_{3}(T)&=&\Lambda-\frac{a_{0}^{2}\lambda}{H_{0}^{2}\tau_{0}^{3}}\left[\frac{513}{8}
    +\frac{3645\sqrt{6}}{16\tau_{0}\sqrt{-T}}-\frac{6561}{4\tau_{0}^{2}T}\right.\\
    &&\left.-\frac{10935\sqrt{6}}{16\tau_{0}^{3}\sqrt{-T^{3}}}\right],\\
    \lim_{t\rightarrow t_{f}}\omega_{3}&=&-1/3,\\
\nonumber V_{3}&=&V_{0}-\frac{a_{0}^{2}\lambda}{H_{0}^{2}\tau_{0}^{3}}\left[\frac{513}{128\pi}
    +\frac{945/\tau_{0}}{64\pi}e^{-\sqrt{2/3}(\varphi-\varphi_{0})}\right.\\
\nonumber  &&+\left.\frac{405/\tau_{0}^{2}}{32\pi}
    e^{-2\sqrt{2/3}(\varphi-\varphi_{0})}+\frac{45/\tau_{0}^{3}}{16\pi}e^{-3\sqrt{2/3}(\varphi-\varphi_{0})}\right].\\
\end{eqnarray}
The evolution of the EoS for this case has been studied in this work in Subsection \ref{Sec4.2}. The interest in this case is motivated by the study of the so-called ``tracker field" when assuming the inflation ends at $\omega_{3} \rightarrow -1/3$ \cite{ZWS98}. The potential $V_{3}$ reduces to $V_{0}$ when $\lambda$ vanishes, its nonvanshing values show general behaviours similar to $V_{1}$. Whereas, Figure \ref{Fig2}\subref{fig2d} shows that, in particular for large negative $\lambda$, the false vacuum is separated by a broad barrier. However, the top of the barrier is quite flat. The decay of the false vacuum is followed by slow-roll inflation allowing a tunneling event from the high energy false vacuum. We find this model fulfills the requirements of \cite{BHS13} to perform well fitting both $E$-mode and $B$-mode polarizations.
\subsubsection{The $V_{n \geq 4}$-models}
For the $V_{4}$-model we find it similar to $V_{2}$-model but with strong flat plateau at $\varphi > 0$ of the false vacuum. At $\varphi < 0$ the symmetry of negative and positive large values of $\lambda$ no longer valid, see plots of Figure \ref{Fig2}\subref{fig2e}. The $V_{n > 4}$-models, the symmetry of large $\pm \lambda$ holds again but they alter their roles, see plots of Figure \ref{Fig2}\subref{fig2f}-\subref{fig2h}. Moreover, the asymptotic EoS goes to increase by $2/9$ with a radiation limit for $\omega_{6}$ while $\omega_{9}$ gives stiff matter, all higher order expansions give unknown matter, so far, with $\omega > 1$. In order to check the consistency of the results which are obtained by the $f(T)$ gravity and by the scalar field of the FRW model, we use equations (\ref{kin})-(\ref{sc_pot}) to evaluate the dynamical EoS of the scalar field, $\omega_{\varphi}=p_{\varphi}/\rho_{\varphi}$, according to the order of expansion. The calculations show that the EoS of the $n=0$ model predict a cosmological constant like DE $\omega(\varphi)=-1$. While the additional terms of the series of $e^{-n\sqrt{2/3}(\varphi-\varphi_{0})}$ in (\ref{sc_pot}) diminishes the effective potential gradually allowing the kinetic energy to showup effectively as the order of expansion increases so that the EoS approaches different asymptotic values with arithmetic sequence $\{-\frac{7}{9}, -\frac{5}{9}, -\frac{1}{3}, ...\}$ respectively. So the rushing of the EoS $\omega_{\varphi}$ towards $\omega_{\varphi}>-1$ is powered by the kinetic energy of the scalar field. When the higher orders of expansion of (\ref{sc_pot}) are taken into consideration, the kinetic energy is enough to end the cosmic inflation allowing the EoS to crossover $\omega_{\varphi}=0$ to enter a matter dominant universe epoch. It is clear that the scalar field analysis is in agreement with the results of the $f(T)$ treatment in Subsection \ref{Sec4.2}.

Now back to the plots of Figure \ref{Fig2}, the overall picture shows similar behaviours for all the models at $\varphi >0$ while they interpolate between Starobinsky and polynomial potentials at $\varphi <0$ with different details as mentioned above. It is known that the $B$-mode polarization excludes the small tensor-to-scalar ratio models such as Starobinsky model \cite{BICEP2}. In contrast, the Planck data restricts the tensor-to-scalar ratio to be small so it excludes inflationary models such as large-field inflation models with a single monomial term \cite{Pl1,Pl2}. In this theory, we find that the inflationary potential interpolates between these two different classes of inflation. These results lead us to investigate the inflationary parameters within this theory.
\section{Single-scalar Field with Double Slowly-rolling Solutions}\label{Sec6}
Assuming that the inflation epoch is dominated by the scalar field potential only. The slow-roll models define two parameters as
\begin{equation}\label{slow_roll}
    \epsilon=\frac{1}{16\pi}\left(\frac{V'}{V}\right)^{2},\qquad \eta=\frac{1}{8\pi}\left(\frac{V''}{V}\right).
\end{equation}
These parameters are called slow-roll parameters. Consequently, the slow-roll inflation is valid where $\epsilon \ll 1$ and $|\eta| \ll 1$ when the potential is dominating. While the end of inflation is characterized by $\textmd{Max}(\epsilon,|\eta|)=1$ as the kinetic term contribution becomes more effective. The slow-roll parameters define two observable parameters
\begin{equation}\label{r_n}
r=16\epsilon,\quad n_{s}=1-6\epsilon+2\eta,
\end{equation}
where $r$ and $n_{s}$ are called the tensor-to-scalar ratio and scalar tilt, respectively. Recent observations by Planck and BICEP2 measure almost the same scalar tilt parameters $n_{s}\sim 0.96$. However, Planck puts an upper limit $r<0.11$ which supports models with small $r$. While BICEP2 sets a lower limit on the $r>0.2$ which supports inflationary models with large $r$. We devote this Section to investigate the capability of the slow-roll models to perform both Planck and BICEP2.
\subsection{Construct a potential from Planck and BICEP2}\label{Sec6.1}
It is clear that Planck and BICEP2 observations agree on the scalar tilt parameter value $n_{s}\sim 0.963$, while they give different tensor-to-scalar ratios $r$. In order to construct a scalar potential performing Planck and BICEP2 data, we found that if the slow-roll parameters (\ref{slow_roll}) satisfy the proportionality relation $\epsilon\propto \eta^{2}$, this gives a chance to find two values of $\eta$ performing the same $n_{s}$ but two different values of $\epsilon$. Consequently, two values of $r$. This can be achieved as follows: using (\ref{r_n}) and the proportionality relation we have
\begin{eqnarray}\label{epseta}
\nonumber n_{s}&=&1-6(\epsilon=\kappa\eta^{2})+2\eta,\\
\textmd{i.e.} \quad \eta^{\pm} &=&\frac{1}{6\kappa}\left(1\pm\sqrt{1+6\kappa(1-n_{s})}\right),\label{quad_eta}
\end{eqnarray}
where $\kappa$ is a constant coefficient and $\eta^{\pm}$ are due to the $\pm$ discriminant. It is clear that there are possibly two different values of the parameter $\eta$ for a single scalar tilt  parameter $n_{s}$. Accordingly, we have $\epsilon=\kappa\eta^{2}=16/r$ which provides double values of $r$ for each $n_{s}$. More concretely, assuming the scalar tilt parameter $n_{s}=0.963$ \cite{Pl2} and substituting into (\ref{epseta}), for a particular choice of the constant $\kappa \sim 30$; then we calculate two possible values of $\eta$ as
\begin{itemize}
\item [(i)] The first solution $\eta^{+}$ has a positive value of $\sim 2.09\times10^{-2}$ which gives $\epsilon^{+} \sim 1.31\times10^{-2}$.
\item [(ii)] The second solution of $\eta^{-}$ has a negative value of $\eta \sim -9.34\times10^{-3}$ which leads to $\epsilon^{-} \sim 3.05\times10^{-3}$.
\end{itemize}
Surely both positive and negative values of $\eta$ give the same scalar tilt $n_{s} \sim 0.963$. Nevertheless, we can get two simultaneous tensor-to-scalar ratios: the first is $r^{+} \sim 0.21$, while the other is smaller $r^{-} \sim 4.9\times10^{-2}$. We conclude that the slow-roll inflationary models which are characterized by the proportionality $\epsilon \propto \eta^{2}$ can perform both $E$-mode and $B$-mode polarizations \cite{BHS13}, when a negative value of $\eta$ is observed near the peak of $\varphi$, it would need to be offset by a positive value of $\eta$ at some later time over a comparable field range in order to get $\epsilon$ to be small again during the period of observable inflation. Generally, at low values of $\kappa$ the model predicts one small value of $r$ in addition to another higher value as required by the $B$-mode polarization inflationary models. Interestingly, at large values of $\kappa$ the model predicts a single value of the tensor-to-scalar parameter $r^{\pm}\rightarrow 0.0987$ which agrees with the upper Plank limit $r_{0.002}<0.11$ at $95\%$ CL.

Moreover, we can investigate the potential pattern which is characterized by the proportionality relation $\epsilon=\kappa\eta^{2}$. Recalling (\ref{slow_roll}), this relation provides a simple differential equation with a solution
\begin{equation}\label{Planck_BICEP2}
V(\varphi)=A+Be^{\pm2\sqrt{\frac{\pi}{\kappa}}\varphi},
\end{equation}
where $A$ and $B$ are constants of integration. In this way, we found that Starobinsky model might be reconstructed naturally from observations if we want our model to perform $E$-mode and $B$-mode polarizations.
\subsection{The slow-roll parameters of the model}\label{Sec6.2}
\begin{figure}
\centering
\includegraphics[scale=.3]{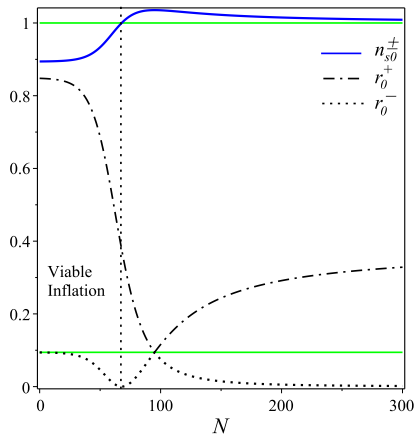}
\caption{The slow-roll parameters vs the $e$-folding number of $V_{0}$-model, where $\Lambda=10^{-30}~s^{-2}$.}
\label{Fig3}
\end{figure}

We calculate the slow roll parameters of the $V_{0}$-model to investigate its capability to predict a vaiable inflation. It can be shown that the $V_{0}$ potential (\ref{V0}) coincides with the potential constructed from Plank and BICEP2 observations (\ref{Planck_BICEP2}). Using (\ref{V0}) and (\ref{slow_roll}), we evaluate the slow-roll parameters
\begin{eqnarray}
  \epsilon_{0} &=& \frac{24 \pi}{\left(12 \pi+\Lambda e^{-2\sqrt{2/3}(\varphi-\varphi_{0})}\right)^{2}},\label{eps}\\
  \eta_{0} &=& \frac{4}{\left(12 \pi+\Lambda e^{-2\sqrt{2/3}(\varphi-\varphi_{0})}\right)}.\label{eta}
\end{eqnarray}
From the slow-roll parameters (\ref{eps}) and (\ref{eta}) of the $V_{0}$-model, it can be shown that the model satisfies the proportionality $\epsilon=\kappa \eta^{2}$, where the proportionality constant $\kappa=\frac{3}{2}\pi$. This relation is not only independent of the values of $\Lambda$ but also it allows a vanishing cosmological constant to exist without affecting the generality of the proportionality relation. Similar relation has been obtained in the literature when studying the leading term behaviour of the Starobinsky inflation as a special case of the $T$-models \cite{CGP14}. The number $e$-folds from the end of inflation to the time of horizon crossing for observable scales
\begin{eqnarray}\label{efold}
\nonumber    N_{*}(\varphi)&=&-8\pi\int_{\varphi}^{\varphi_{f}}{\frac{V}{V'}}d\varphi\\
   &=&-2\sqrt{6}\pi\left(\varphi-\frac{\Lambda}{16\pi}\sqrt{\frac{2}{3}}e^{-2\sqrt{\frac{2}{3}}(\varphi-\varphi_{f})}\right),
\end{eqnarray}
where $\varphi_{f}$ is the value of $\varphi$ at the end of inflation. Using (\ref{eps})-(\ref{efold}) we can reexpress the slow-roll parameters as functions of $N$. At the end of inflation, i.e. Max($\epsilon$, $|\eta|$)=1, for the allowed range of $N_{*}\sim 60$ the observable cosmological wavelengths exit the Hubble radius for a minimum and maximum inflationary scale we have $\varphi_{f}=40$. In the following steps we show the capability of the $\epsilon \propto \eta^{2}$ models to predict two tensor-to-scalar ratios for a single spectral tilt at different $e$-folds. Using (\ref{efold}) we evaluate the spectral scalar tilt $n_{s0}$ of the $V_{0}$-model as a function of $N$. Substituting $n_{s0}(N)$ into (\ref{quad_eta}) we evaluate two values $\eta_{0}^{+}$ and $\eta_{0}^{-}$, for each we have two possible values $\epsilon_{0}^{+}$ and $\epsilon_{0}^{-}$, respectively. Finally, we have two tensor-to-scalar ratios $r_{0}^{\pm}=16\epsilon_{0}^{\pm}$, while the two expected spectral scalar tilt are identical, i.e. $n_{s0}^{+}=n_{s0}^{-}$. This procedure enables to draw the plots of Figure \ref{Fig3}.  The plots represent the evolution of the observable parameters vs the $e$-folding number. As is clear the spectral scalar tilt has a single pattern represented by $n_{s0}^{\pm}$ plot, while the tensor-to-scalar ratio has two distinguished patterns represented by $r_{0}^{+}$ and $r_{0}^{-}$ plots. The $r_{0}^{+}$ has a lower limit $>0.094$ which is in agreement with the $B$-mode polarization models. On the other hand, the $r_{0}^{-}$ plot has an upper limit $<0.094$ which is in agreement with the $E$-mode polarization models.

In Figure \ref{Fig3}, the viable inflation is characterized by $n_{s0}^{\pm}<1$ region. The region with $n_{s0}^{\pm}$ exceeding unity is Max$(n_{s0}^{\pm})<1.04$, although this small deviation from the scale invariance of the scalar power spectrum is acceptable at the level of theoretical predictions, it can be excluded as it appears beyond the acceptable range of the $e$-folds $N>60$. Interestingly, at large $e$-folding number the spectral scalar tilt goes again below unity so that $(1-n_{s0}^{\pm})\sim 4\times 10^{-10}$. We highlight some results in Table \ref{T1} showing that the $V_{0}$-model can predict large tensor-to-scalar ratios of the $B$-mode polarization as well as small ratios as required by the $E$-mode polarization.
\begin{table}
\caption{The predicted parameters of the $V_{0}$-model}
\label{T1}
\begin{tabular*}{\columnwidth}{@{\extracolsep{\fill}}lrrcc@{}}
\hline
$N$                & $59\qquad$            & $66\qquad$            & $99\quad$             & $\infty$\\
\hline
$\eta_{0}^{+}$     & $8.48\times 10^{-2}$  & $7.37\times 10^{-2}$  & $3.85\times 10^{-2}$  & $7.07\times 10^{-2}$\\[6pt]
$\eta_{0}^{-}$     & $-1.40\times 10^{-2}$ & $-3.00\times 10^{-3}$ & $3.23\times 10^{-2}$  & $0\quad$\\[6pt]
$\epsilon_{0}^{+}$ & $3.39\times 10^{-2}$  & $2.56\times 10^{-2}$  & $6.98\times 10^{-3}$  & $2.36\times 10^{-2}$\\[6pt]
$\epsilon_{0}^{-}$ & $9.27\times 10^{-4}$  & $4.25\times 10^{-5}$  & $4.90\times 10^{-3}$  & $0$\\[6pt]
$n_{s0}^{\pm}$     & $0.966\qquad$         & $0.994\qquad$         & $1.04\quad$           & $\preceq 1$\\[6pt]
$r_{0}^{+}$        & $0.542\qquad$         & $0.410\qquad$         & $0.112\quad$          & $0.377$\\[6pt]
$r_{0}^{-}$        & $0.015\qquad$         & $6.81\times 10^{-4}$  & $7.85\times 10^{-2}$  & $0$\\
\hline
\end{tabular*}
\end{table}

Similar procedure can be applied to the $V_{1}$-model where the $\epsilon=\kappa \eta^{2}$ relation still valid with larger proportionality constant $\kappa=6\pi$. Accordingly, the tensor-to-scalar ratios can be achieved, e.g. at $N=55$ we have ($r_{1}^{-}$, $r_{1}^{+}$, $n_{s1}^{\pm}$) = ($3.23\times 10^{-2}$, $0.237$, $0.967$) where the scalar tilt in this case always $n_{s1}^{\pm}<1$. Also, this procedure can be used to put constraints on the choice of conformal weight in the power law and Starobinsky inflation models from the CMB observations\footnote{This work is in progress now.}.
\section{Discussions and Final Remarks}\label{Sec7}
We investigate a special class of $f(T)$ gravity governed by FLU assumptions. In this model, we assume the vanishing of the coefficient of the sectional curvature in the modified Friedmann equations. This enables to identify a particular class of $f_{n}(T)\propto \sum \frac{1}{\sqrt{-T^{n}}}$ gravity, usually hidden by the SFU assumption, cannot be covered by TEGR theory. The FLU model, at large Hubble regime, is consistent with the cosmic inflation scenario. Also, it enables to evaluate a dynamical EoS of the cosmic fluid.  In a particular case $n=3$, the fluid can evolve from phantom initial phase crossing the phantom divided line ($\omega=-1$) to radiation and possibly stiff-matter with a quintessence fate. As a matter of fact, the dynamical EoS avoids the usual problems of the cosmological constant DE models. In addition, there is no need to assume a large cosmological constant at early universe.

We provide an alternative approach to study $f(T)$ inflation by considering the case when the torsion is made of a scalar field $\varphi$. This approach allows the $f(T)$ to predict the inflationary observable parameters. We show that the teleparallel torsion scalar can couple to the kinetic energy of the scalar field, while its potential $V(\varphi)$ can be induced by the $f(T)$ gravity. In this case, we define a scalar field sensitive to the spacetime symmetry with a potential induced exactly from extended teleparallel gravity. This leads finally to a gravitational quintessence model governed by Friedmann and Klein-Gordon equations.

The obtained quintessence model covers different inflationary models according to the expansion limit. In this study, we give more attention to the $V_{0}$-model. It has been shown that the $V_{0}$ model relates the slow-roll parameters by the proportionality $\epsilon=\kappa \eta^{2}$ which in agreement with the model constructed from Planck and BICEP2 data. The model can be classified as power law inflation which has been shown that is capable of performing two tensor-to-scalar ratios consistent with both $E$ and $B$ modes of polarization, while it predicted scalar spectral index still unique. In particular, the observable parameters evolution with the $e$-folding $N$ can be shown in Figure \ref{Fig3} and Table \ref{T1}. We just highlighted some results of the other models, while further studies are left for future work. Higher orders potentials interpolate between Starobinsky and quadratic like inflations.

Finally, we would like to mention that at large $\kappa$ the model predicts a single tensor-to-scalar parameter $r\rightarrow 0.0987$ in agreement with Planck observations limit. This result can be used to constrain the choice of the conformal weight in power law and Starobinsky models to the observable parameters of the cosmic inflation epoch.
\subsection*{Acknowledgments}
The authors would like to thank the anonymous Referee for her/his valuable comments which, indeed, help to improve the work. This article is partially supported by the Egyptian Ministry of Scientific Research under project No. 24-2-12.
\bibliographystyle{epjc}
\bibliography{1409.7199}

\begin{thebibliography}{10}
\providecommand{\url}[1]{\texttt{#1}}
\providecommand{\urlprefix}{URL }
\providecommand{\eprint}[2][]{\url{#2}}

\bibitem{SN98}
A.~G. {Riess}, A.~V. {Filippenko}, P.~C. {et. al.}, Astrophys. J. \textbf{116},
  1009 (1998), \eprint{astro-ph/9805201}

\bibitem{AP2013}
R.~Aldrovandi, J.~G. Pereira, \emph{Teleparallel Gravity: An Introduction},
  vol. 173 of \emph{Fundamental Theories of Physics} (Springer, Springer
  Dordrecht Heidelberg New York London, 2013)

\bibitem{1010.1041}
B.~{Li}, T.~P. {Sotiriou}, J.~D. {Barrow}, Phys. Rev. D \textbf{83}, 6, 064035
  (2011), \eprint{1010.1041}

\bibitem{1012.4039}
T.~P. {Sotiriou}, B.~{Li}, J.~D. {Barrow}, Phys. Rev. D \textbf{83}, 10, 104030
  (2011), \eprint{1012.4039}

\bibitem{E28}
A.~{Einstein}, \emph{Riemann-Geometrie mit Aufrechterhaltung des Begriffes des
  Fernparallelismus}, 217 -- 221, Phys.-math. Klasse (Preussische Akademie der
  Wissenschaften, Sitzungsberichte, 1928)

\bibitem{M52}
F.~I. {Mikhail}, \emph{Relativistic cosmology and some related problems in
  general relativity}, Ph. {D.} {Thesis}, University of London (1952)

\bibitem{MM56}
W.~H. {McCrea}, F.~I. {Mikhail}, Royal Society of London Proceedings Series A
  \textbf{235}, 11 (1956)

\bibitem{M64}
F.~{Mikhail}, Al Nuovo Cimento \textbf{series X}, 32, 886 (1964)

\bibitem{MW77}
F.~I. {Mikhail}, M.~I. {Wanas}, Royal Society of London Proceedings Series A
  \textbf{356}, 471 (1977)

\bibitem{M78}
C.~{M{\o}ller}, K. Dan. Vidensk. Selsk. Mat. Fys. Skr. \textbf{89}, 13 (1978)

\bibitem{HS79}
K.~{Hayashi}, T.~{Shirafuji}, Phys. Rev. D \textbf{19}, 3524 (1979)

\bibitem{U56}
R.~{Utiyama}, Physical Rev. \textbf{101}, 5, 1597 (1956)

\bibitem{K61}
T.~W.~B. {Kibble}, J. Math. Phys. \textbf{2}, 2, 212 (1961)

\bibitem{S64}
D.~W. {Sciama}, Rev. Mod. Phys. 463--469 (1964)

\bibitem{H76}
F.~W. {Hehl}, P.~{von der Heyde}, G.~D. {Kerlick}, et~al., Reviews of Modern
  Physics \textbf{48}, 393 (1976)

\bibitem{U80}
R.~{Utiyama}, Progress of Theoretical Physics \textbf{64}, 2207 (1980)

\bibitem{NS07}
N.~L. {Youssef}, A.~M. {Sid-Ahmed}, Reports on Mathematical Physics
  \textbf{60}, 39 (2007), \eprint{0604111}

\bibitem{NS13}
N.~L. {Youssef}, W.~A. {Elsayed}, Reports on Mathematical Physics \textbf{72},
  1 (2013), \eprint{1209.1379}

\bibitem{W09}
M.~I. {Wanas}, Modern Physics Letters A \textbf{24}, 1749 (2009),
  \eprint{0801.1132}

\bibitem{WK11}
M.~I. {Wanas}, M.~M. {Kamal}, Modern Physics Letters A \textbf{26}, 2065
  (2011), \eprint{1103.4121}

\bibitem{YST12}
N.~L. {Youssef}, A.~M. {Sid-Ahmed}, E.~H. {Taha}, Int. J. Geom. Meth. Mod.
  Phys. \textbf{10}, 7, 1350029 (2013), \eprint{1206.4505}

\bibitem{FF07}
R.~{Ferraro}, F.~{Fiorini}, Phys. Rev. D \textbf{75}, 8, 084031 (2007),
  \eprint{gr-qc/0610067}

\bibitem{FF08}
R.~{Ferraro}, F.~{Fiorini}, Phys. Rev. D \textbf{78}, 12, 124019 (2008),
  \eprint{gr-qc/0812.1981}

\bibitem{BF09}
G.~R. {Bengochea}, R.~{Ferraro}, Phys. Rev. D \textbf{79}, 12, 124019 (2009),
  \eprint{0812.1205}

\bibitem{L10}
E.~V. {Linder}, Phys. Rev. D \textbf{81}, 12, 127301 (2010), \eprint{1005.3039}

\bibitem{1008.4036}
K.~{Bamba}, C.-Q. {Geng}, C.-C. {Lee}, ArXiv e-prints  (2010),
  \eprint{1008.4036}

\bibitem{1011.0508}
K.~{Bamba}, C.-Q. {Geng}, C.-C. {Lee}, et~al., JCAP \textbf{1}, 021 (2011),
  \eprint{1011.0508}

\bibitem{Y2011}
R.-J. {Yang}, EPL (Europhysics Letters) \textbf{93}, 60001 (2011),
  \eprint{1010.1376}

\bibitem{CDDS11}
S.-H. {Chen}, J.~B. {Dent}, S.~{Dutta}, et~al., Phys. Rev. D \textbf{83}, 2,
  023508 (2011), \eprint{1008.1250}

\bibitem{FF011}
R.~{Ferraro}, F.~{Fiorini}, Phys. Rev. D \textbf{84}, 8, 083518 (2011),
  \eprint{1109.4209}

\bibitem{FF11}
R.~{Ferraro}, F.~{Fiorini}, Phys. Lett. B \textbf{702}, 75 (2011),
  \eprint{1103.0824}

\bibitem{IS12}
L.~{Iorio}, E.~N. {Saridakis}, Mon. Not. R. Astron. Soc. \textbf{427}, 1555
  (2012), \eprint{1203.5781}

\bibitem{CGS13}
S.~{Capozziello}, P.~A. {Gonz{\'a}lez}, E.~N. {Saridakis}, et~al., JHEP
  \textbf{2}, 39 (2013), \eprint{1210.1098}

\bibitem{Nashed1}
G.~G.~L. {Nashed}, Phys. Rev. D \textbf{88}, 10, 104034 (2013),
  \eprint{1311.3131}

\bibitem{Nashed2}
G.~G.~L. {Nashed}, General Relativity and Gravitation \textbf{45}, 1887 (2013),
  \eprint{1502.05219}

\bibitem{RHTMM13}
M.~E. {Rodrigues}, M.~J.~S. {Houndjo}, J.~{Tossa}, et~al., JCAP \textbf{11},
  024 (2013), \eprint{1306.2280}

\bibitem{Nashed3}
G.~G.~L. {Nashed}, EPL (Europhysics Letters) \textbf{105}, 10001 (2014),
  \eprint{1501.00974}

\bibitem{BFG15}
C.~{Bejarano}, R.~{Ferraro}, M.~J. {Guzm{\'a}n}, Eur. Phys. J. C \textbf{75},
  77 (2015), \eprint{1412.0641}

\bibitem{Nashed4}
G.~G.~L. {Nashed}, Journal of the Physical Society of Japan \textbf{84}, 4,
  044006 (2015)

\bibitem{Nashed5}
G.~G.~L. {Nashed}, International Journal of Modern Physics D \textbf{24},
  1550007 (2015)

\bibitem{1205.3421}
K.~{Bamba}, S.~{Capozziello}, S.~{Nojiri}, et~al., Astrophysics and Space
  Science \textbf{342}, 155 (2012), \eprint{1205.3421}

\bibitem{BNO14}
K.~{Bamba}, S.~{Nojiri}, S.~D. {Odintsov}, Phys. Lett. 257--264 (2014),
  \eprint{1401.7378}

\bibitem{BO14}
K.~{Bamba}, S.~D. {Odintsov}, ArXiv: 1402.7114  (2014), \eprint{1402.7114}

\bibitem{JMM14}
M.~{Jamil}, D.~{Momeni}, R.~{Myrzakulov}, International Journal of Theoretical
  Physics  (2014), \eprint{1309.3269}

\bibitem{HLOS14}
T.~{Harko}, F.~S.~N. {Lobo}, G.~{Otalora}, et~al., Phys. Rev. D \textbf{89},
  12, 124036 (2014), \eprint{1404.6212}

\bibitem{NH14}
G.~G.~L. {Nashed}, W.~{El Hanafy}, European Physical Journal C \textbf{74}, 10,
  3099 (2014), arXiv: 1403.0913, \eprint{1403.0913}

\bibitem{WH14}
M.~I. {Wanas}, H.~A. {Hassan}, International Journal of Theoretical Physics
  \textbf{53}, 3901 (2014)

\bibitem{HN14}
W.~{El Hanafy}, G.~G.~L. {Nashed}  (2014), \eprint{1410.2467}

\bibitem{1503.05281}
Y.~{Wu}, Z.-C. {Chen}, J.~{Wang}, et~al., ArXiv e-prints  (2015),
  \eprint{1503.05281}

\bibitem{1503.07427}
E.~L.~B. {Junior}, M.~E. {Rodrigues}, M.~J.~S. {Houndjo}, ArXiv e-prints
  (2015), \eprint{1503.07427}

\bibitem{CCDDS11}
Y.-F. {Cai}, S.-H. {Chen}, J.~B. {Dent}, et~al., Classical and Quantum Gravity
  \textbf{28}, 21, 215011 (2011), \eprint{1104.4349}

\bibitem{CQSW14}
Y.-F. {Cai}, J.~{Quintin}, E.~N. {Saridakis}, et~al., JCAP \textbf{7}, 033
  (2014), \eprint{1404.4364}

\bibitem{NHS14}
G.~{Nashed}, W.~{El Hanafy}, S.~{Ibrahim}  (2014), \eprint{1411.3293}

\bibitem{KS114}
G.~{Kofinas}, E.~N. {Saridakis}, Phys. Rev. D \textbf{90}, 8, 084044 (2014),
  \eprint{1404.2249}

\bibitem{KS214}
G.~{Kofinas}, E.~N. {Saridakis}, Phys. Rev. D \textbf{90}, 8, 084045 (2014),
  \eprint{1408.0107}

\bibitem{KS314}
G.~{Kofinas}, G.~{Leon}, E.~N. {Saridakis}, Classical and Quantum Gravity
  \textbf{31}, 17, 175011 (2014), \eprint{1404.7100}

\bibitem{Pl2}
{Planck Collaboration}, P.~A.~R. {Ade}, N.~{Aghanim}, et~al., Astronomy \&
  Astrophysics \textbf{571} (2014), \eprint{1303.5082}

\bibitem{Pl1}
{Planck Collaboration}, P.~A.~R. {Ade}, N.~{Aghanim}, et~al., Astronomy \&
  Astrophysics \textbf{571} (2014), \eprint{1303.5076}

\bibitem{BICEP2}
{BICEP2 Collaboration}, P.~A.~R. {Ade}, et~al., Physical Review Letters
  \textbf{112}, 24, 241101 (2014), \eprint{1403.3985}

\bibitem{M2013}
J.~W. {Maluf}, Annalen der Physik \textbf{525}, 339 (2013), \eprint{1303.3897}

\bibitem{R32}
H.~P. {Robertson}, Ann. Math. \textbf{33}, 496 (1932)

\bibitem{FMR15}
O.~{Farooq}, D.~{Mania}, B.~{Ratra}, Astrophysics and Space Science
  \textbf{357}, 11 (2015)

\bibitem{DI12}
J.~N. {Dossett}, M.~{Ishak}, Phys. Rev. D \textbf{86}, 10, 103008 (2012),
  \eprint{1205.2422}

\bibitem{ZZCZ14}
J.-F. {Zhang}, M.-M. {Zhao}, J.-L. {Cui}, et~al., European Physical Journal C
  \textbf{74}, 3178 (2014), \eprint{1409.6078}

\bibitem{S02}
V.~{Sahni}, Classical and Quantum Gravity \textbf{19}, 3435 (2002),
  \eprint{astro-ph/0202076}

\bibitem{W2012}
M.~I. {Wanas}, Adv. High Energy Phys. \textbf{2012}, Article ID 752613, 10
  pages (2012)

\bibitem{XSH96}
H.-j. {Xie}, T.~{Shirafuji}  (1996), \eprint{9603006}

\bibitem{R74}
V.~E. {Rochev}, Theoretical and Mathematical Physics \textbf{18}, 160 (1974)

\bibitem{HRR78}
S.~{Hojman}, M.~{Rosenbaum}, M.~P. {Ryan}, et~al., Phys. Rev. D \textbf{17},
  3141 (1978)

\bibitem{H90}
R.~T. {Hammond}, Classical and Quantum Gravity \textbf{7}, 2107 (1990)

\bibitem{HO01}
F.~W. {Hehl}, Y.~N. {Obukhov}, in C.~{L{\"a}mmerzahl}, C.~W.~F. {Everitt},
  F.~W. {Hehl}, eds., \emph{Gyros, Clocks, Interferometers ...: Testing
  Relativistic Gravity in Space}, vol. 562 of \emph{Lecture Notes in Physics,
  Berlin Springer Verlag}, 479 (2001), \eprint{gr-qc/0001010}

\bibitem{AP98}
V.~C. {de Andrade}, J.~G. {Pereira}, General Relativity and Gravitation
  \textbf{30}, 263 (1998), \eprint{9706070}

\bibitem{E55}
A.~{Einstein}, \emph{The meaning of relativity}, 5th edn. (the Princeton
  University Press, 1955)

\bibitem{FRM13}
J.~B. {Fonseca-Neto}, C.~{Romero}, S.~P.~G. {Martinez}, General Relativity and
  Gravitation \textbf{45}, 1579 (2013), \eprint{1211.1557}

\bibitem{NA2005}
L.~Y. {Nabil}, A.~M. {Sid-Ahmed}, in M.~Wanas, ed., \emph{The international
  conference on developing and extending Einstein's ideas}, 19--21, Egyptian
  Relativity Group (NART, Cairo, Egypt, 2005)

\bibitem{KLR214}
R.~{Kallosh}, A.~{Linde}, D.~{Roest}, JHEP \textbf{8}, 52 (2014),
  \eprint{1405.3646}

\bibitem{St80}
A.~A. {Starobinsky}, Phys. Lett. B \textbf{91}, 99 (1980)

\bibitem{ZWS98}
I.~{Zlatev}, L.~{Wang}, P.~J. {Steinhardt}, Physical Review Letters
  \textbf{82}, 896 (1999), \eprint{astro-ph/9807002}

\bibitem{BHS13}
R.~{Bousso}, D.~{Harlow}, L.~{Senatore}, Phys.Rev. D \textbf{91}, 8, 083527
  (2015), \eprint{1309.4060}

\bibitem{CGP14}
Y.-F. {Cai}, J.-O. {Gong}, S.~{Pi}, Phys. Lett. B \textbf{738}, 20 (2014),
  \eprint{1404.2560}

\end{thebibliography}
\end{document}